%% file: main.tex
\documentclass[journal]{IEEEtran}
\IEEEoverridecommandlockouts
\makeatletter
\def\endthebibliography{%
  \def\@noitemerr{\@latex@warning{Empty `thebibliography' environment}}%
  \endlist
}
\makeatother

\usepackage[most]{tcolorbox}
\usepackage{xcolor}     

\usepackage{xcolor}        

\usepackage{cite}
\usepackage{amsmath,amssymb,amsfonts}
\usepackage{algorithm}
\usepackage{algpseudocode}
\usepackage{graphicx} 
\usepackage{bmpsize}
\usepackage{grffile}
\usepackage{subfigure}
\usepackage{subcaption}
\usepackage[belowskip=-10pt,aboveskip=1pt]{caption}
\usepackage{caption}
\usepackage{textcomp}
\usepackage{float}
\usepackage{stfloats}
\usepackage{xcolor}
\usepackage{balance}
\usepackage{dsfont}
\usepackage{multirow}
\usepackage{ntheorem}
\usepackage[colorlinks,
            linkcolor=black,
            anchorcolor=black,
            citecolor=black]{hyperref}
\usepackage{amsmath}
\pdfoutput=1

\ifodd 0
 
\newcommand{\com}[1]{{\color{blue}#1}} 

\else

\newcommand{\com}[1]{}
\fi

\def\BibTeX{{\rm B\kern-.05em{\sc i\kern-.025em b}\kern-.08em
    T\kern-.1667em\lower.7ex\hbox{E}\kern-.125emX}}

\begin{document} 
\title{{An LLM-Agent-Based Framework for Age of Information Optimization in Heterogeneous Random Access Networks}\\
}
\author{Fang~Liu*,~Erchao Zhu*,~Jiedan Tan,~Jingwen Tong,~Taotao~Wang~and~Shengli Zhang
\thanks{F. Liu, E. Zhu, J. Tan, J. Tong, T.~Wang and S. Zhang are with the College of Electronics and Information Engineering, Shenzhen University, Shenzhen, China, e-mails: \{liuf, eejwentong, ttwang, zsl\}@szu.edu.cn; \{2410044083,2510044003\}@mails.szu.edu.cn }
\thanks{F. Liu and E. Zhu contributed equally to this work. (Corresponding author: T. Wang and J. Tong)}}

\maketitle
\begin{abstract}
With the rapid expansion of the Internet of Things (IoT) and heterogeneous wireless networks, the Age of Information (AoI) has emerged as a critical metric for evaluating the performance of real-time and personalized systems. While AoI-based random access is essential for next-generation applications such as the low-altitude economy and indoor service robots, existing strategies, ranging from rule-based protocols to learning-based methods, face critical challenges, including idealized model assumptions, slow convergence, and poor generalization. In this article, we propose \emph{Reflex-Core}, a novel Large Language Model (LLM) agent-based framework for AoI-driven random access in heterogeneous networks. 
By devising an ``Observe-Reflect-Decide-Execute'' closed-loop mechanism, this framework integrates Supervised Fine-Tuning (SFT) and Proximal Policy Optimization (PPO) to enable optimal, autonomous access control. Based on the Reflex-Core framework, we develop a Reflexive Multiple Access (RMA) protocol and a priority-based RMA variant for intelligent access control under different heterogeneous network settings. Experimental results demonstrate that in the investigated scenarios, the RMA protocol achieves up to a $\mathrm{14.9}\%$ reduction in average AoI compared with existing baselines, while the priority-based version improves the convergence rate by approximately $\mathrm{20}\%$.
\end{abstract}


\section{Introduction}\label{SecIntro}
\input{Introduction}

\input{MainPart}

\bibliographystyle{./IEEEtran}
\bibliography{./IEEEabrv}
\end{document}

%% file: Introduction.tex
{T}{he} rapid evolution of the Internet of Things (IoT) has led to a paradigm shift in wireless connectivity, characterized by a massive density of devices and an increasingly complex landscape of heterogeneous networks~\cite{da2014internet}. In these environments, diverse wireless technologies and protocols, ranging from scheduled access like TDMA to contention-based mechanisms like ALOHA, must coexist and compete for limited spectral and temporal resources. Random access remains the cornerstone of these networks, providing a flexible and decentralized means for nodes to initiate communication~\cite{yu2019deep}. However, as next-generation networks expand into emerging applications such as the low-altitude economy and indoor service robotics, the traditional focus on throughput and latency is no longer sufficient to meet the stringent demands of real-time coordination.

Recently, the Age of Information (AoI) has emerged as a critical metric for evaluating the performance of real-time and personalized systems. Unlike traditional delay metrics, AoI measures the ``freshness'' of information from the perspective of the destination, making it indispensable for applications where timely status updates are vital for decision-making~\cite{tong2022age}. Despite its importance, optimizing AoI-enabled random access in heterogeneous networks faces significant challenges. Existing strategies, which span from rigid rule-based protocols to modern learning-based methods, often suffer from idealized model assumptions that fail in real-world deployments~\cite{chiariotti2021spectrum}. Furthermore, Deep Reinforcement Learning (DRL) approaches~\cite{xu2020aoi, zhao2022deep, leng2019age} frequently encounter slow convergence rates and a lack of generalization, requiring extensive retraining whenever the network topology or traffic patterns shift. These limitations highlight the need for new methodologies that can efficiently and autonomously optimize the AoI-based random access in heterogeneous networks.

Recent advances in LLMs and LLM-based Agents have demonstrated remarkable capabilities in reasoning, planning, and tool invocation~\cite{tong2025wirelessagent}. These agents possess the potential to address the complexities of AoI-based random access by translating raw network statistics into semantic insights, allowing for more flexible and human-like decision-making. By treating protocol optimization as a reasoning task, \cite{kwon2025cp} proposes CP-AgentNet, which can successfully utilize generative agents for protocol design in heterogeneous environments.
However, early attempts to integrate LLMs into communication protocols have been limited; they often rely on general-purpose prompts that lack the specialized optimization objectives required for information freshness and fail to provide a robust mechanism for continuous policy improvement.

To overcome these limitations, we propose \textit{Reflex-Core}, a novel LLM-based framework designed for AoI-based random access in heterogeneous networks. This framework is established on an ``Observe-Reflect-Decide-Execute" closed-loop mechanism, which enables nodes to analyze their performance and refine their behavior through semantic reasoning~\cite{yao2022react}. A key innovation of Reflex-Core is its integration of Supervised Fine-Tuning (SFT) and Proximal Policy Optimization (PPO). By fine-tuning the underlying LLM with an AoI-centric reward signal and agent experiences, the agent’s ``reflections'' can translate into optimal, autonomous transmission decisions. Based on this framework, we devise the Reflexive Multiple Access (RMA) protocol, which empowers nodes to coexist intelligently with legacy protocols while minimizing system-wide AoI.

Furthermore, recognizing that different applications require varying levels of information freshness, we introduce a priority-based  RMA protocol. This mechanism incorporates node-specific priorities directly into the LLM-based priority understanding and decision-making process, allowing for intelligent differentiation of access opportunities. By leveraging the  coordination capabilities of the Reflex-Core framework, the priority-based RMA protocol ensures that critical nodes (e.g., control links in a low-altitude economy) receive preferential treatment without starving lower-priority updates. Our experimental results indicate that this priority-aware approach not only maintains superior information freshness but also achieves a $20\%$ faster convergence rate compared with state-of-the-art baselines, marking a significant step toward self-evolving, intelligent wireless networks.
In summary, this work makes the following contributions. 
\begin{itemize}
\item  We introduce Reflex-Core, a novel LLM-based framework for AoI optimization using an ``Observe-Reflect-Decide-Execute'' mechanism, enabling interpretable, semantic-driven random access in heterogeneous networks.
\item We integrate SFT and PPO to align LLM reasoning with AoI-centric  objectives, enabling the agent to autonomously refine its strategies and optimize information freshness without manual intervention.
\item 
We develop the RMA protocol and a priority-based RMA protocol, providing intelligent, differentiated access control enabled by semantic priority reasoning and adaptive coordination.
\item Experimental results show that RMA reduces average AoI by up to $14.9\%$ and achieves a $20\%$ faster convergence rate compared to state-of-the-art baselines across diverse heterogeneous scenarios. 
\end{itemize}

This paper is structured as follows. Section II presents the related work, covering AoI optimization and LLM-based agent systems in wireless communications. Section III introduces the system model, including the overall heterogeneous network scenario and AoI definitions. Section IV details the system design, describing the Reflex-Core framework. Section V introduces the RMA protocol. Section VI provides numerical results through extensive experiments. Finally, Section VII concludes this paper.

%% file: MainPart.tex
\section{Related Work}
\subsection{AoI Optimization}
AoI has emerged as a cornerstone metric for quantifying status freshness in real-time systems \cite{kaul2012status, yates2021age}. Initial research focused on homogeneous networks, with theoretical frameworks deriving average AoI expressions for ALOHA-based systems \cite{yang2021understanding} and investigating the trade-offs between update intervals and throughput in IoT systems \cite{bae2022age}. To handle the stochastic nature of wireless channels, RL has been widely adopted. For instance, researchers have modeled AoI optimization as Markov Decision Processes (MDPs) to balance energy consumption \cite{xu2020aoi}, joint throughput-AoI objectives \cite{zhao2022deep}, and ad hoc network scheduling \cite{leng2019age}.

The rise of heterogeneous wireless networks has introduced multi-dimensional optimization challenges \cite{chiariotti2021spectrum}. Existing studies have addressed heterogeneity in source reliability \cite{gindullina2021age}, traffic types \cite{zhang2024optimal}, and control requirements \cite{ayan2021age}. Further advancements include using fluid limits to analyze varied channel conditions \cite{jiang2021analyzing} and developing AoI-driven cache update algorithms \cite{ma2020age}. However, most existing works assume a unified access protocol across the network, leaving a critical gap: protocol heterogeneity, where nodes using fundamentally different multiple-access mechanisms (e.g., TDMA and ALOHA) contend for shared spectrum.


\subsection{LLM-based Agents in Wireless Networks}
LLM-based autonomous agents have demonstrated transformative potential across diverse domains \cite{lim2024large, bossema2025llm}, supported by foundational architectures for self-reflection \cite{shinn2023reflexion} and policy gradient optimization \cite{yao2023retroformer, wang2024survey}. In wireless communication, research has evolved from domain-specific knowledge alignment \cite{shao2024wirelessllm} and 6G perception modules \cite{xu2024large} to infrastructure-level optimizations like WirelessAgent for dedicated network slicing \cite{tong2025wirelessagent}. Furthermore, LLMs have outperformed traditional CNNs in urban coverage tasks \cite{sevim2024large} and enabled autonomous throughput protocol design through multi-agent collaboration~\cite{kwon2025cp}.

Recent efforts have begun to intersect LLMs with AoI optimization, primarily utilizing them as auxiliary operators within evolutionary algorithms for UAV routing \cite{wei2025laura} and V2I access control \cite{xu2025enhanced}, or as predictive models for time-series forecasting \cite{picano2025performance}. However, these approaches largely relegate LLMs to offline optimization or forecasting roles rather than utilizing them for real-time, autonomous decision-making.

To the best of our knowledge, Reflex-Core is the first framework to directly integrate LLM-based reflection and fine-tuning (SFT/PPO) into a real-time random access protocol. By shifting from auxiliary tool use to a closed-loop ``Observe-Reflect-Decide-Execute'' architecture, our work provides a novel, interpretable, and self-evolving solution specifically tailored for AoI-centric access control in heterogeneous networks.

\section{System Model}
\subsection{Network Scenarios}
\begin{figure}[!t]
    \centering
    \includegraphics[width=1.0\linewidth]{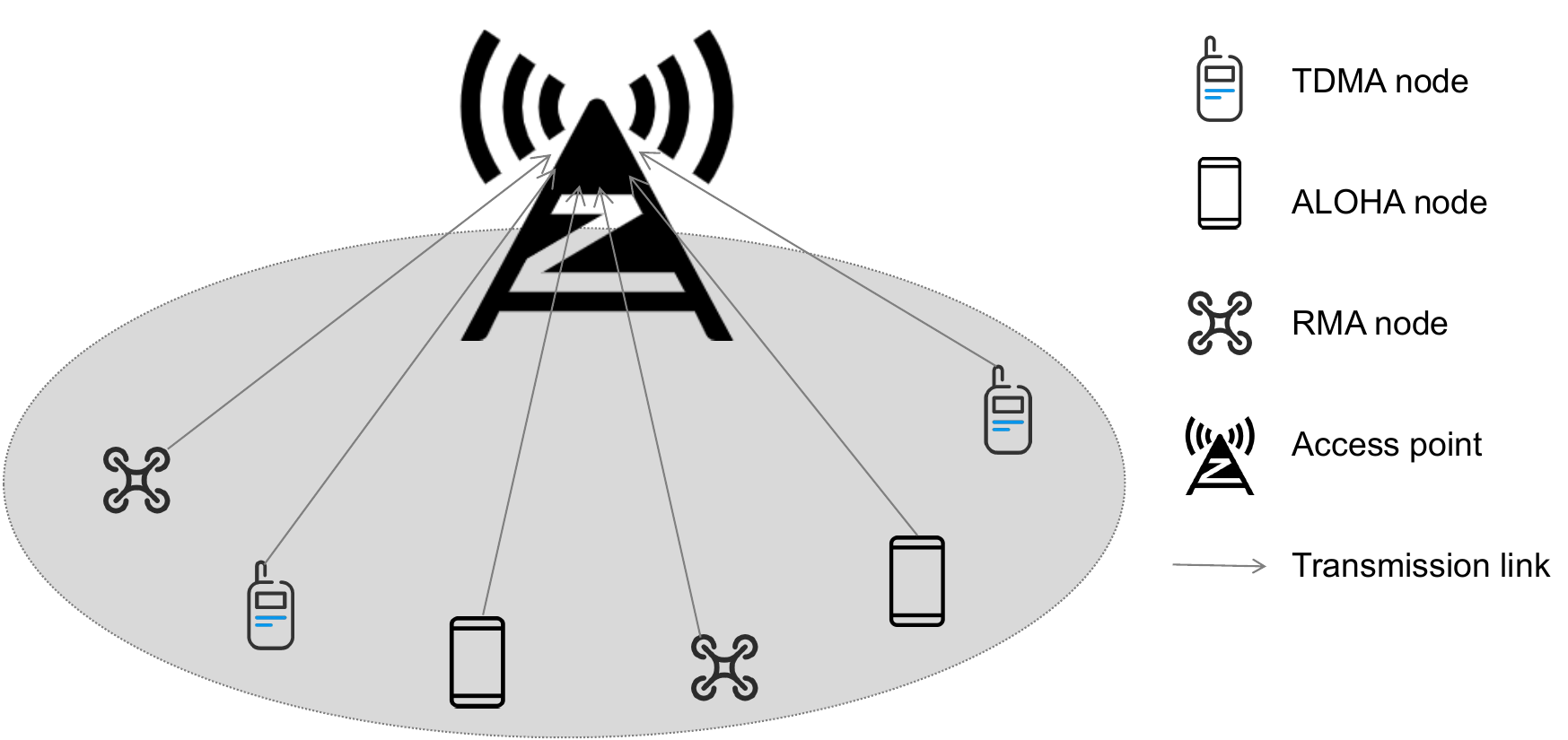}
    \caption{An illustration of the heterogeneous random access network.}
    \label{fig:aoi_network}
\end{figure}

We consider a time-slotted heterogeneous wireless network as illustrated in Fig.~\ref{fig:aoi_network}, comprising multiple nodes that transmit status updates to a common Access Point (AP) via a shared wireless channel. Time is discretized into fixed-length slots. We adopt the ``generate-at-will'' model, assuming that each node always has a fresh data packet available for transmission at the beginning of every time slot. Transmissions are synchronized; nodes may only initiate transmission at the start of a slot, and the process must conclude within that same slot. If two or more nodes transmit simultaneously, a collision occurs, resulting in the failure of all involved transmissions. The AP periodically broadcasts feedback to the nodes, detailing network status metrics such as transmission outcomes, collision statistics, and the system performance indicator, which in this work is the AoI of the nodes.

The network comprises nodes utilizing distinct access protocols: TDMA, ALOHA, and the proposed Reflexive Multiple Access (RMA). The specific characteristics of each node type are detailed below:
\begin{itemize}
    \item TDMA: Each TDMA node operates on a fixed schedule, transmitting in pre-assigned time slots within a periodic frame structure. The slot allocation remains consistent across consecutive frames, ensuring deterministic access behavior.

    \item  ALOHA: Each ALOHA node employs a randomized access strategy, independently attempting transmission in any given time slot with a fixed probability~$q$.

    \item  RMA: Each RMA node employs the RMA protocol based on the Reflex-Core framework. Through an ``Observe-Reflect-Decide-Execute'' closed-loop process, the node intelligently adapts its transmission probability in each time slot based on environmental observations and accumulated historical insights.
\end{itemize}

\subsection{Definition of AoI}
In this heterogeneous wireless network, nodes send timestamped data packets to the AP through a shared wireless channel. The timestamp of each data packet is set to the start time of its transmission slot to reflect information freshness.

We denote the instantaneous AoI of node $i$ in time slot $n$ by $\delta_i(n)$, which is defined as
\begin{equation}
\delta_i(n) = n - \sigma_i(n),
\end{equation}
where $\sigma_i(n)$ represents the timestamp of the latest data packet from node $i$ successfully received by the AP up to time slot~$n$.

\begin{figure}[!t]
    \centering
    \includegraphics[width=0.8\linewidth]{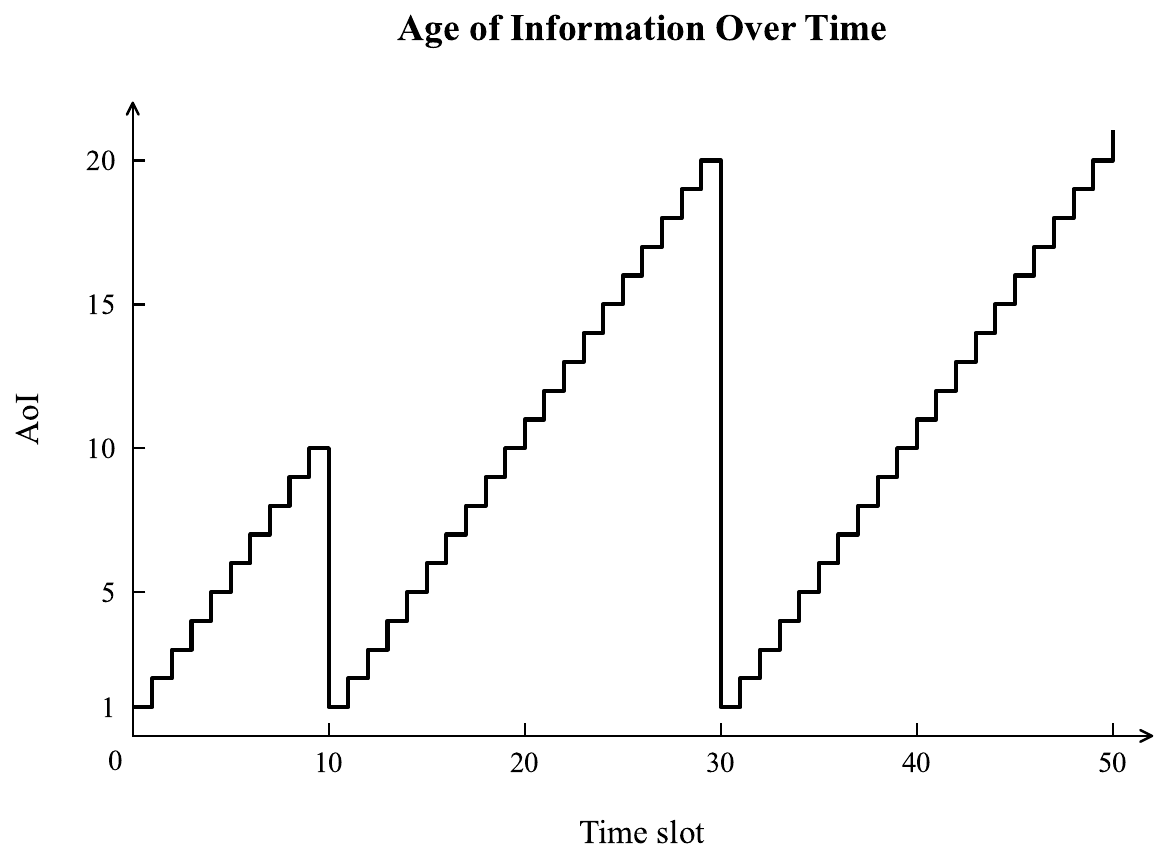}
    \caption{Definition of the instantaneous AoI of a node over time.}
    \label{fig:aoi_definition}
\end{figure}

As shown in Fig.~\ref{fig:aoi_definition}, the instantaneous AoI of each node grows linearly over time and drops sharply to 1 when successful transmission events occur, intuitively reflecting the definition in Equation (1).

Based on the definition of instantaneous AoI, the average AoI of node $i$ at time $n$ can be defined as
\begin{equation}
\Delta_i(n) = \frac{1}{n} \sum_{\ell=1}^{n} \delta_i(\ell).
\end{equation}
While the system's average AoI (System AoI) is defined as the cumulative average AoI over all $m$ nodes at slot $n$
  \begin{equation}
  \Delta_{\text{sys}}(n) = \sum_{i=1}^{m} \Delta_i(n) = \sum_{i=1}^{m} \frac{1}{n} \sum_{\ell=1}^{n} \delta_i(\ell).
  \end{equation}

The objective of the RMA protocol is to enhance the AoI performance through agent-based intelligent access control. By continuously refining their transmission strategies, RMA nodes aim to optimize information freshness while ensuring efficient coexistence with nodes operating under different protocols in heterogeneous networks. To achieve this goal, we propose Reflex-Core, a novel LLM-based framework for intelligent access control.

\section{The Reflex-Core Framework}
\begin{figure}[!t]
    \centering
    \includegraphics[width=0.5\textwidth]{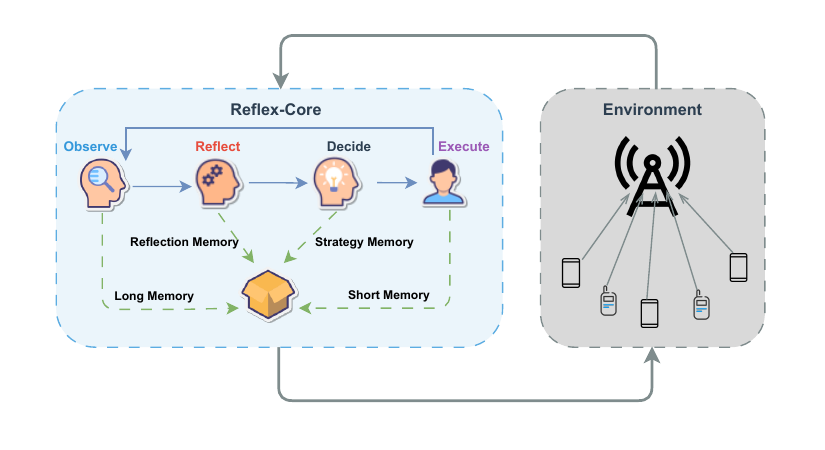}
    \caption{An illustration of the Reflex-Core framework, which contains the main part and environment part. The main part consists of an ``Observe-Reflect-Decide-Execute" closed-loop mechanism.}
    \label{fig:framework}
\end{figure}

The Reflex-Core framework, illustrated in Fig.~\ref{fig:framework}, facilitates intelligent random access control through a synergistic interaction between the agent core and the network environment. The framework architecture comprises five primary modules: Observe, Reflect, Decide, Execute, and Memory. These components operate in a ``Observe-Reflect-Decide-Execute'' closed-loop mechanism to achieve autonomous policy optimization.
In the following, we first introduce the five modules separately in Section~\ref{SubSec_CM}. Then, we employ the SFT and PPO technologies to improve the domain capabilities of the foundation model in Section~\ref{SubSec_FTM}.

\subsection{The Closed-Loop Decision Mechanism }\label{SubSec_CM}

\begin{figure}[!t]
      \centering
      \includegraphics[width=0.99\linewidth]{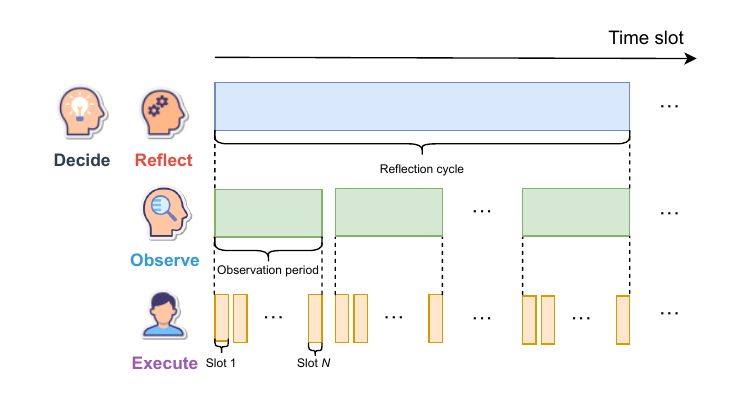}
      \caption{An illustration of the hierarchical temporal structure in Reflex-Core, showing the relationship between reflection cycles, observation periods, and time slots.}
      \label{fig:temporal_structure}
\end{figure}

To harmonize the high-frequency network dynamics with the low-frequency reasoning of the LLM, Reflex-Core employs a hierarchical temporal structure, as illustrated in Fig.~\ref{fig:temporal_structure}. This structure operates on three distinct time scales, each governed by a specific module:

\begin{itemize}
    \item Fine-grained Execution (Slot Level): Driven by the Execute Module, the network operates in continuous time slots. In each slot, the module executes real-time transmission decisions based on the current policy to maintain low-level responsiveness.
    \item Aggregated Observation (Period Level): Managed by the Observe Module, the system filters high-frequency noise by aggregating raw metrics every $N$ slots into an \textit{Observation Period}. This transforms discrete signals into stable statistical states.
    \item Asynchronous Reflection (Cycle Level): Orchestrated by the Decide and Reflect Modules, the agent accumulates context over $O$ observation periods to form a \textit{Reflection Cycle}. It triggers reflection only at this macro-scale to optimize the global strategy, ensuring computational efficiency.
\end{itemize}


The specific functions of the five constituent modules within this temporal framework are detailed below.

\subsubsection{Observe Module}

The Observe module functions as a telemetry and analysis engine that periodically generates semantic feedback from raw network trajectories. Every $N$ time slots, it extracts key performance indicators, including AoI gradients, success rates, and state distributions (e.g., idle, success, or collision). Based on a comparative analysis of current and historical windows, the module generates the perturbation term
\[
\Delta p_i(o) = f_{\text{obs}}(F_o),
\]
where $F_o$ represents the semantic observation feedback. This feedback is stored in the Long-term Memory to serve as the empirical foundation for subsequent reflection, transforming numerical network states into actionable linguistic insights.

\subsubsection{Reflect Module}
The Reflect module serves as the cognitive core of the framework, utilizing an LLM to perform linguistic reinforcement learning. Inspired by the Reflexion architecture \cite{shinn2023reflexion}, it evaluates the quality of current strategies by  accessing the historical observation statistics stored in the Long-term Memory.  For single-node scenarios, it optimizes system-wide AoI; for multi-node deployments, it focuses on individual node freshness. By analyzing the failure modes and successes documented in the Long-term Memory, the module generates targeted, semantic optimization suggestions.  These linguistic insights and the corresponding context are then stored in the Reflection Memory, allowing the agent to continuously diagnose and rectify suboptimal behaviors through historical self-reference.

\subsubsection{Decide Module}
The Decide module translates the high-level semantic suggestions from the Reflect module into concrete global transmission parameters. It updates the global policy via the relation $p_i(t+1) = p_i(t) + \beta \cdot f(\text{Reflection}(t))$, where $\beta$ denotes the learning rate and $f(\cdot)$ represents the mapping of linguistic advice to numerical adjustments. Beyond policy updates, this module manages the Strategy Memory, a repository of historically superior strategies. By retrieving successful past configurations, the Decide module ensures that the framework maintains a steady performance trajectory and avoids catastrophic forgetting during the iterative optimization process.

\subsubsection{Execute Module}
The Execute module handles real-time transmission control by executing local dynamic adjustments to the transmission probability in each time slot. It synthesizes the global policy $p_i(t)$ provided by the Decide module with a local perturbation term $\Delta p_i(o)$ from the Observer module. For a node $i$ at slot $n$, the final transmission probability is defined as $p^{\text{final}}_{i}(n) = p_i(t) + \Delta p_i(o)$. This mechanism ensures that while the agent adheres to a long-term strategy, it remains responsive to transient channel variations and collision patterns, effectively bridging high-level semantic reasoning with low-level network execution. Following the transmission attempt, the specific execution outcomes and instantaneous network states are immediately recorded in the Short-term Memory, providing the raw trajectory data required for the subsequent observation phase.

\subsubsection{Memory Module}
The Memory module serves as the foundational knowledge base of Reflex-Core, systematically archiving multi-scale data to drive the ``Observe-Reflect-Decide-Execute'' loop. It is structured into four specialized components: 
\begin{itemize}
    \item \textbf{Short-term Memory ($M_{\text{short}}$):} Records state-action pairs $\{(s(n), a(n))\}_{n=1}^{N}$ for each slot $n$ within an observation period, providing raw data for the Observe module.
    \item \textbf{Long-term Memory ($M_{\text{long}}$):} Aggregates semantic observation statistics $\{O_k\}_{k=1}^{O}$ across an entire reflection cycle, forming the empirical basis for the Reflect module's reasoning.
    \item \textbf{Reflection Memory ($M_{\text{reflection}}$):} Maintains a historical log of linguistic reflections, AoI improvements, and system contexts $\{R_r, A_r, C_r\}_{r=1}^{T}$ to track the evolution of the agent's cognitive insights.
    \item \textbf{Strategy Memory ($M_{\text{strategy}}$):} Archives successful global strategies that achieved verified AoI reductions, enabling the Decide module to reference or inherit high-performing policies.
\end{itemize} 

This hierarchical architecture ensures that the framework can leverage both transient environmental feedback and long-term strategic experience for continuous optimization. However, the general-purpose foundation models often lack the domain-specific precision required to navigate the complex trade-offs of wireless random access. To bridge this gap and enhance the agent's reasoning accuracy within communication contexts, we implement a specialized fine-tuning mechanism. By leveraging SFT to establish a baseline of expert-like protocol analysis and PPO to align linguistic reflections with objective AoI performance metrics, we transform the LLM from a general reasoner into a high-performance network controller capable of generating nuanced, effective transmission strategies.

\subsection{The SFT and PPO-Based Post-Training}\label{SubSec_FTM}
To enhance the domain-specific reasoning and adaptability of the Reflect module, we implement a structured fine-tuning protocol. This mechanism utilizes policy gradient optimization to align the LLM's linguistic outputs with objective network performance metrics, facilitating the autonomous evolution of strategy generation. The process is organized into four sequential stages.

\subsubsection{Data Collection and Construction}
During system operation, the Reflect module generates dual reflection candidates for each trigger. We record the complete context, including long-term memory and historical reflections, alongside the resulting LLM outputs. The efficacy of each candidate is quantified by the subsequent AoI reduction. Schemes yielding superior AoI performance are labeled as ``high-quality,'' while others are designated ``low-quality''. This comparative dataset forms the empirical basis for preference learning and reward modeling.

\subsubsection{Supervised Fine-Tuning (SFT) Stage}
We first perform SFT on the base Reflect LLM using a curated corpus of logically coherent and high-performing reflection samples. This stage aligns the model's linguistic structure and reasoning logic with the specific requirements of AoI optimization. The SFT objective is defined as: 
\begin{equation} 
\mathcal{L}_{\text{SFT}} = -\sum_{i=1}^{N} \log P_\theta(y_i | x_i), \end{equation} 
where $(x_i, y_i)$ denotes the input-reflection pairs and $\theta$ represents the model parameters.

\subsubsection{Reward Model Training Stage}
A reward model is trained to distinguish between reflection qualities by using AoI reduction as an implicit reinforcement signal. The reward $r_\phi$ is mapped to the observed performance gain: For a single RMA node: 
\begin{equation} 
r_\phi(\text{Reflection}) = \Delta{\text{sys}}^{\text{before}} - \Delta{\text{sys}}^{\text{after}}. 
\end{equation} 
For multiple RMA nodes: 
\begin{equation} 
r_\phi(\text{Reflection}) = \Delta_{i}^{\text{before}} - \Delta_{i}^{\text{after}}, 
\end{equation} 
where $\phi$ is the reward model parameter. This ensures that reflections leading to improved information freshness receive positive reinforcement, effectively calibrating the model's judgment of ``quality''.

\subsubsection{PPO Fine-Tuning Stage}
Finally, we employ the PPO algorithm \cite{schulman2017proximal} to optimize the Reflect LLM module. The objective function utilizes a clipping mechanism to maintain training stability: 
\begin{equation} \mathcal{L}_{\text{PPO}} = \mathbb{E}\left[\min\left(r_t(\theta) \hat{A}_t, \text{clip}(r_t(\theta), 1-\epsilon, 1+\epsilon) \hat{A}_t\right)\right], \end{equation} 
where the importance weight $r_t(\theta)$ is defined as: 
\begin{equation} r_t(\theta) = \frac{\pi_\theta(a_t|s_t)}{\pi_{\theta_{\text{old}}}(a_t|s_t)}. 
\end{equation} Here, $\hat{A}_t$ represents the advantage estimate and $\epsilon$ is the clipping hyperparameter. This stage constrains the update magnitude, ensuring the model converges toward generating high-quality, targeted reflections that consistently minimize system AoI through self-evolutionary optimization.

\section{The Proposed Protocols}\label{SubSec_RMA}
In the following, we first introduce the RMA protocol built on the Reflex-Core framework. Then, we present the priority-aware RMA protocol, considering more practical applications.

\subsection {The Reflexive Multiple Access Protocol}
In fact, the RMA protocol can be viewed as an application of the Reflex-Core framework designed to achieve near-optimal information freshness in heterogeneous wireless environments. In a typical deployment, nodes utilizing the RMA protocol coexist with legacy nodes, such as those governed by TDMA or ALOHA. The protocol operates by treating the complex interactions of these diverse access mechanisms as environmental observations. By leveraging the ``Observe-Reflect-Decide-Execute'' loop, RMA nodes do not require prior knowledge of the coexisting protocols; instead, they autonomously infer the latent scheduling patterns of the network through semantic reasoning. This allows the RMA protocol to dynamically exploit available spectrum gaps and avoid collisions, effectively providing a self-adaptive coordinating layer that minimizes the system-wide AoI without manual reconfiguration.

To implement RMA in a random access network, each intelligent node is equipped with a local instance of the Reflex-Core modules, operating over a time-slotted physical layer. During the execution phase, the node samples its transmission decision based on the probability $p_i^{\text{final}}(n)$, which is continuously refined by the higher-level cognitive modules. Implementation is facilitated by a hierarchical control structure: at the slot level, the Decide module performs low-latency execution, while at the observation and reflection levels, the agent processes accumulated telemetry to update the global strategy. This architecture allows the protocol to be integrated into existing software-defined radio (SDR) platforms or IoT gateways, where the LLM-based reflection can be offloaded to an edge server or executed locally using lightweight, fine-tuned models. Consequently, the RMA protocol transforms the traditional random access problem into a continuous learning task, ensuring robust performance across varying network densities and traffic loads.

While the standard RMA protocol excels in fully distributed scenarios through autonomous inference, certain mission-critical applications, such as industrial automation or emergency response, require explicit service differentiation to prioritize urgent data streams. To address this, we extend the framework into a centralized coordination mechanism where the relative importance of different information sources is explicitly recognized. By moving from a blind distributed approach to a priority-aware centralized architecture, RMA nodes gain visibility into the priority hierarchy of their peers. This shared knowledge enables the Reflex-Core framework to transition from simple collision avoidance to sophisticated, priority-driven scheduling. In this mode, the Reflex-Core framework leverages cross-node priority information to orchestrate transmission strategies that guarantee ultra-low AoI for high-priority nodes while maintaining efficient resource utilization for the rest of the network.

\subsection{The Priority-Aware RMA Protocol} 
A fundamental advantage of the Reflex-Core framework is the LLM's inherent capability to derive node priorities from semantic context, i.e., a cognitive function that traditional numerical optimization lacks. By interpreting natural language descriptors, the LLM qualitatively distinguishes between traffic types based on their mission-criticality and AoI sensitivity. For example, in an autonomous vehicular network, the model intuitively identifies that safety-critical alerts, such as collision avoidance signals, require immediate channel access to maintain system stability. Conversely, it recognizes that non-urgent data, such as infotainment updates, can tolerate higher AoI. This semantic-driven approach replaces rigid, pre-defined mathematical weights with a dynamic utility-based assessment, allowing for a more flexible and context-aware allocation of wireless resources.

To operationalize this capability in multi-node environments, we introduce a centralized priority-sharing mechanism within the RMA protocol. In this configuration, each RMA node’s priority status is disseminated across the framework. This shared awareness facilitates differentiated scheduling through three integrated stages: 
\begin{enumerate}
\item \textbf{Priority-Informed Observation:} The Observe module weights the impact of collisions and idle slots differently based on the priority of the nodes involved, generating biased perturbation terms $\Delta p_i(o)$ that favor high-priority nodes. Mathematically, the agent sets different initial transmission probabilities for different priority nodes during initialization, affecting the initial global policy generated by the execute module:
\begin{equation}
p_i^{\text{initial}} =
\begin{cases}
p_{\text{HP}}^{\text{initial}}, & \text{if } \text{Priority}(i) = \text{High} \\
p_{\text{LP}}^{\text{initial}}, & \text{if } \text{Priority}(i) = \text{Low}
\end{cases},
\end{equation}
where $p_{\text{HP}}^{\text{initial}} > p_{\text{LP}}^{\text{initial}}$, ensuring high-priority nodes get more transmission opportunities at system startup.

\item \textbf{Strategic Reflection:} The Reflect module utilizes the shared priority hierarchy to perform reasoning, proposing aggressive transmission strategies for critical nodes while suggesting cooperative back-off behaviors for lower-priority ones. In addition, a behavioral perturbation mechanism is used for adjusting the access probability. Perturbation terms generated by the Observe Agent are related to priority settings, with high-priority nodes receiving positive perturbations and low-priority nodes being affected by negative perturbations:
\begin{equation}
\Delta p_i(n) =
\begin{cases}
\theta_{\text{HP}} + \epsilon_i(n), & \text{if } \text{Priority}(i) = \text{High} \\
\theta_{\text{LP}} + \epsilon_i(n), & \text{if } \text{Priority}(i) = \text{Low}
\end{cases},
\end{equation}
where $\theta_{\text{HP}}$ is the positive perturbation for high-priority nodes, $\theta_{\text{LP}}$ is the negative perturbation for low-priority nodes, and $\epsilon_i(n)$ is random exploration noise.

\item \textbf{Coordinated Execution:} The Execute module synthesizes these semantic insights into a globally coherent policy, ensuring that nodes with strict AoI constraints receive prioritized service while maintaining network-wide efficiency. Specifically, the priority information runs through the decision process through semantic understanding. High-priority nodes receive more active adjustment suggestions, while low-priority nodes adopt relatively moderate adjustment ranges. In strategy memory library management, successful strategies for high-priority nodes are stored preferentially. The complete prompt templates and operational examples that implement these priority-based mechanisms are detailed in Appendix B, which demonstrates how priority constraints are systematically embedded in agent decision-making processes through carefully engineered prompts. 
\end{enumerate}

This mechanism transforms random access into a priority-aware coordination task. By converting static constraints into multi-dimensional semantic insights, it enables explicit priority control, ensuring high-priority nodes maintain lower AoI and high-value information remains fresh in heterogeneous environments.
Moreover, this approach offers inherent interpretability, as every strategy adjustment is backed by a transparent semantic rationale. Notably, by leveraging direct LLM reasoning, it bypasses complex mathematical modeling and exhaustive parameter tuning, effectively validating the framework's ability to map intricate network control challenges onto semantic understanding tasks.

  \begin{figure*}[!t]
      \centering

      \begin{minipage}{0.32\textwidth}
          \centering
          \includegraphics[width=\textwidth]{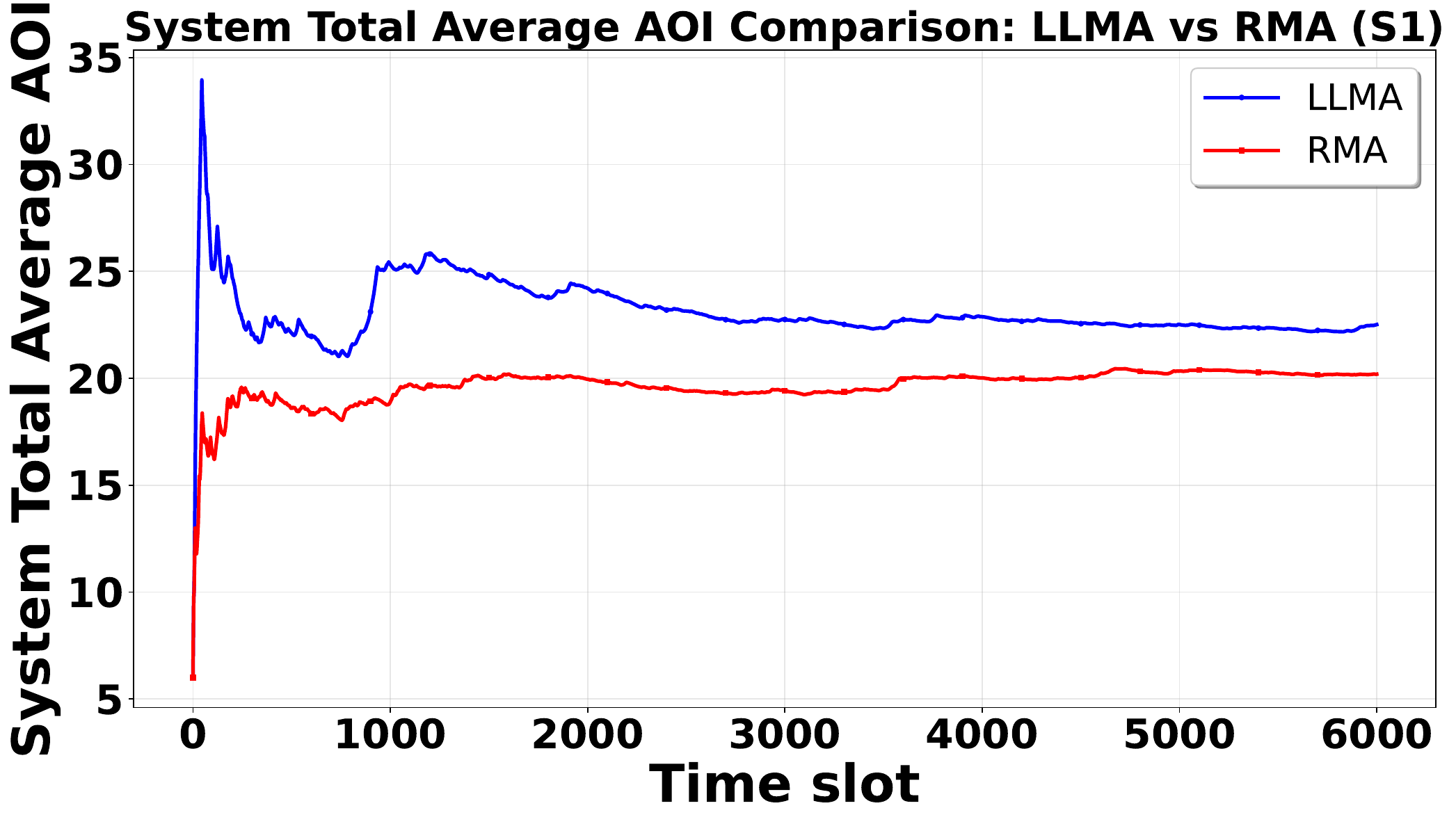}
          \scriptsize{(a) Scenario 1}
      \end{minipage}
      \hfill
      \begin{minipage}{0.32\textwidth}
          \centering
          \includegraphics[width=\textwidth]{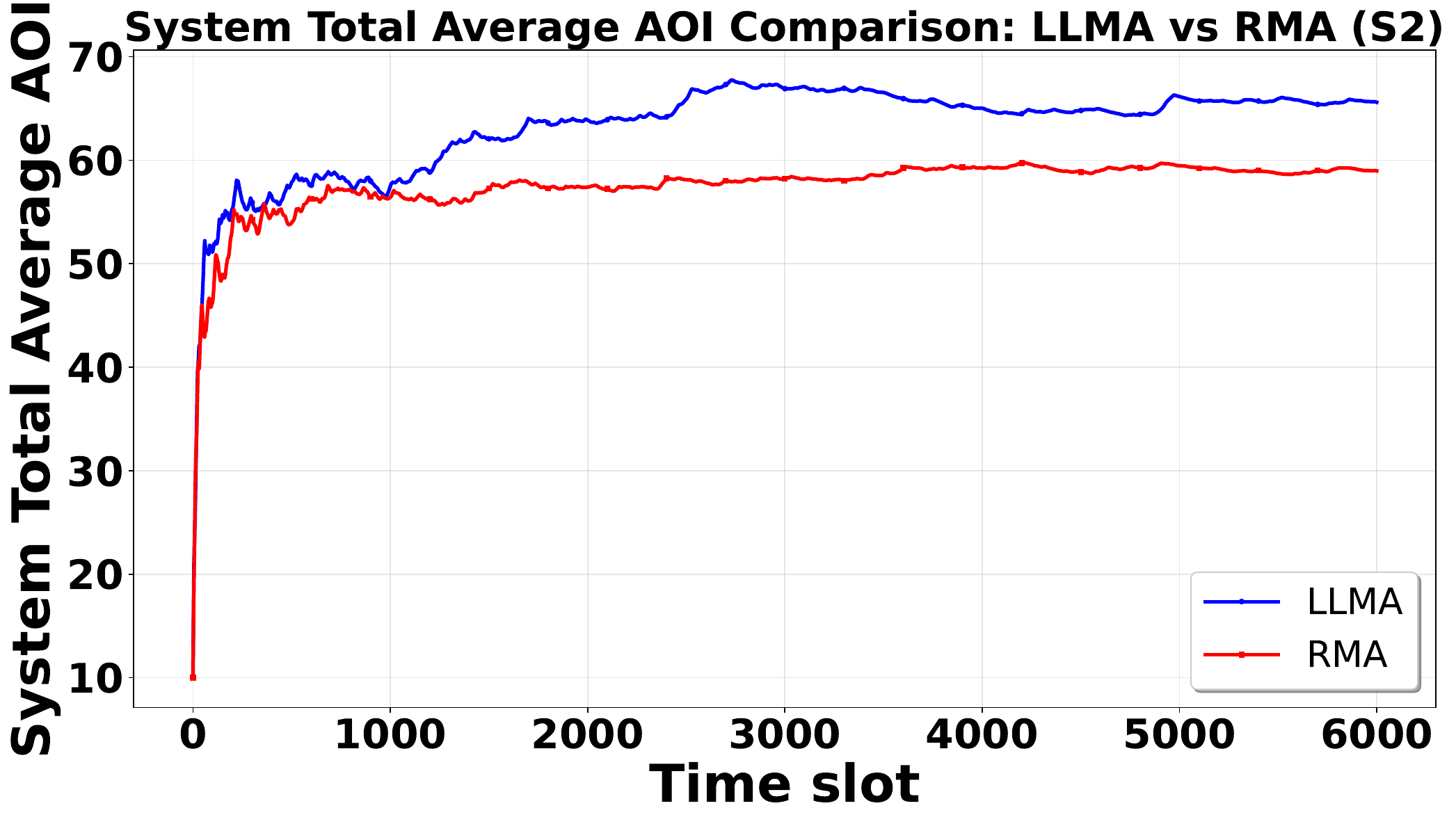}
          \scriptsize{(b) Scenario 2}
      \end{minipage}
      \hfill
      \begin{minipage}{0.32\textwidth}
          \centering
          \includegraphics[width=\textwidth]{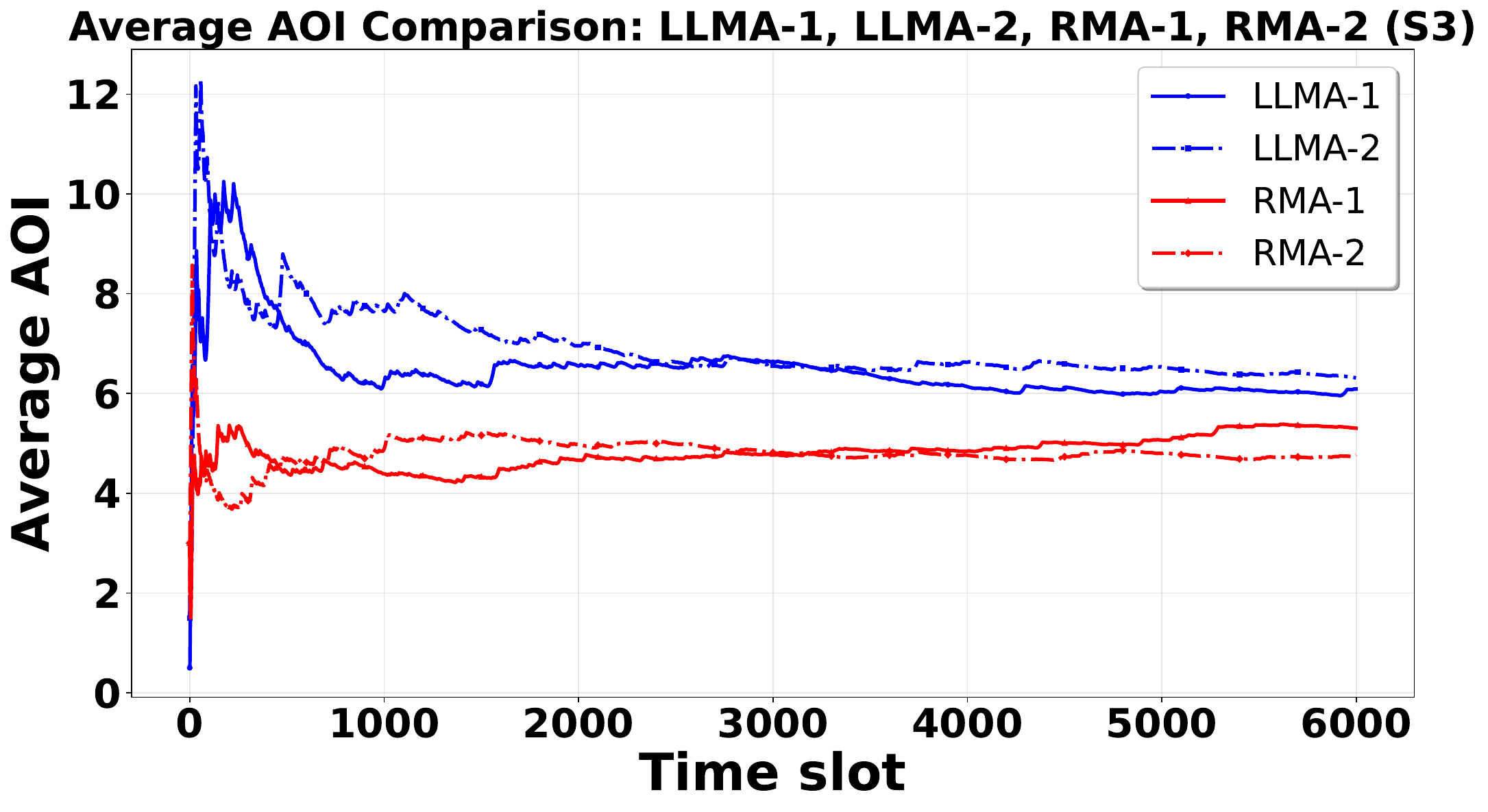}
          \scriptsize{(c) Scenario 3}
      \end{minipage}

      \vspace{0.3cm}

      \begin{minipage}{0.32\textwidth}
          \centering
          \includegraphics[width=\textwidth]{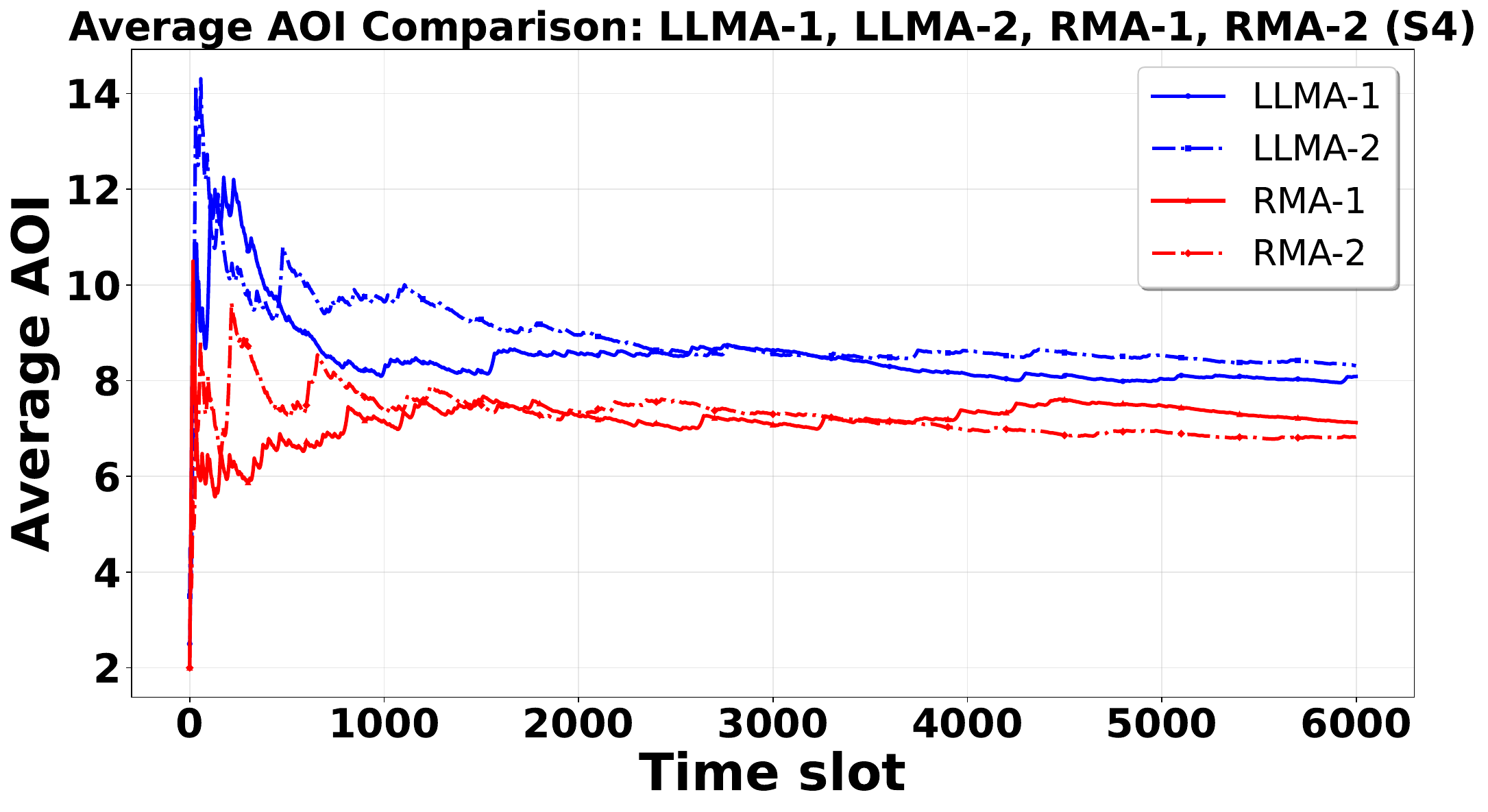}
          \scriptsize{(d) Scenario 4}
      \end{minipage}
      \hfill
      \begin{minipage}{0.32\textwidth}
          \centering
          \includegraphics[width=\textwidth]{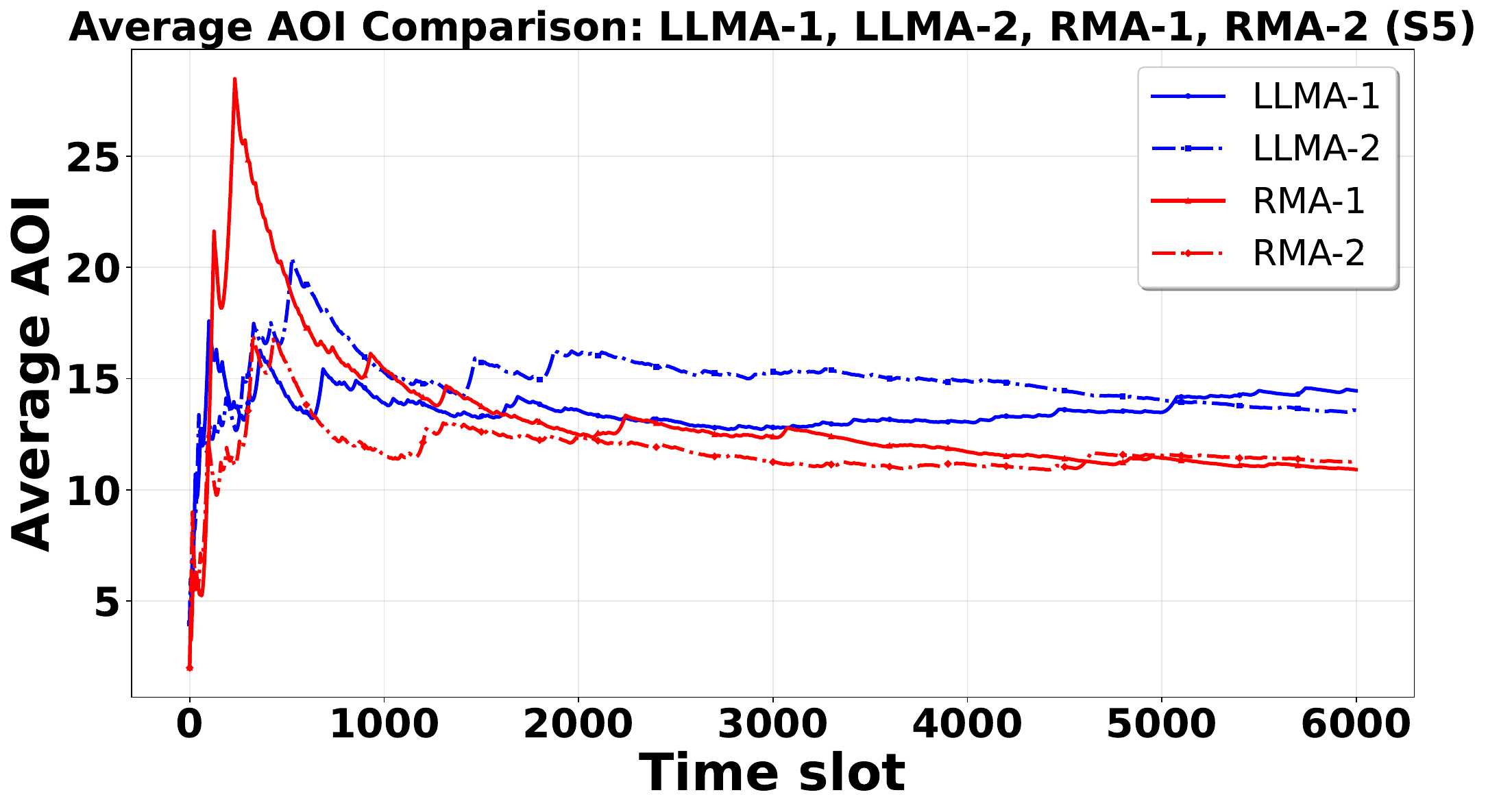}
          \scriptsize{(e) Scenario 5}
      \end{minipage}
      \hfill
      \begin{minipage}{0.32\textwidth}
          \centering
          \includegraphics[width=\textwidth]{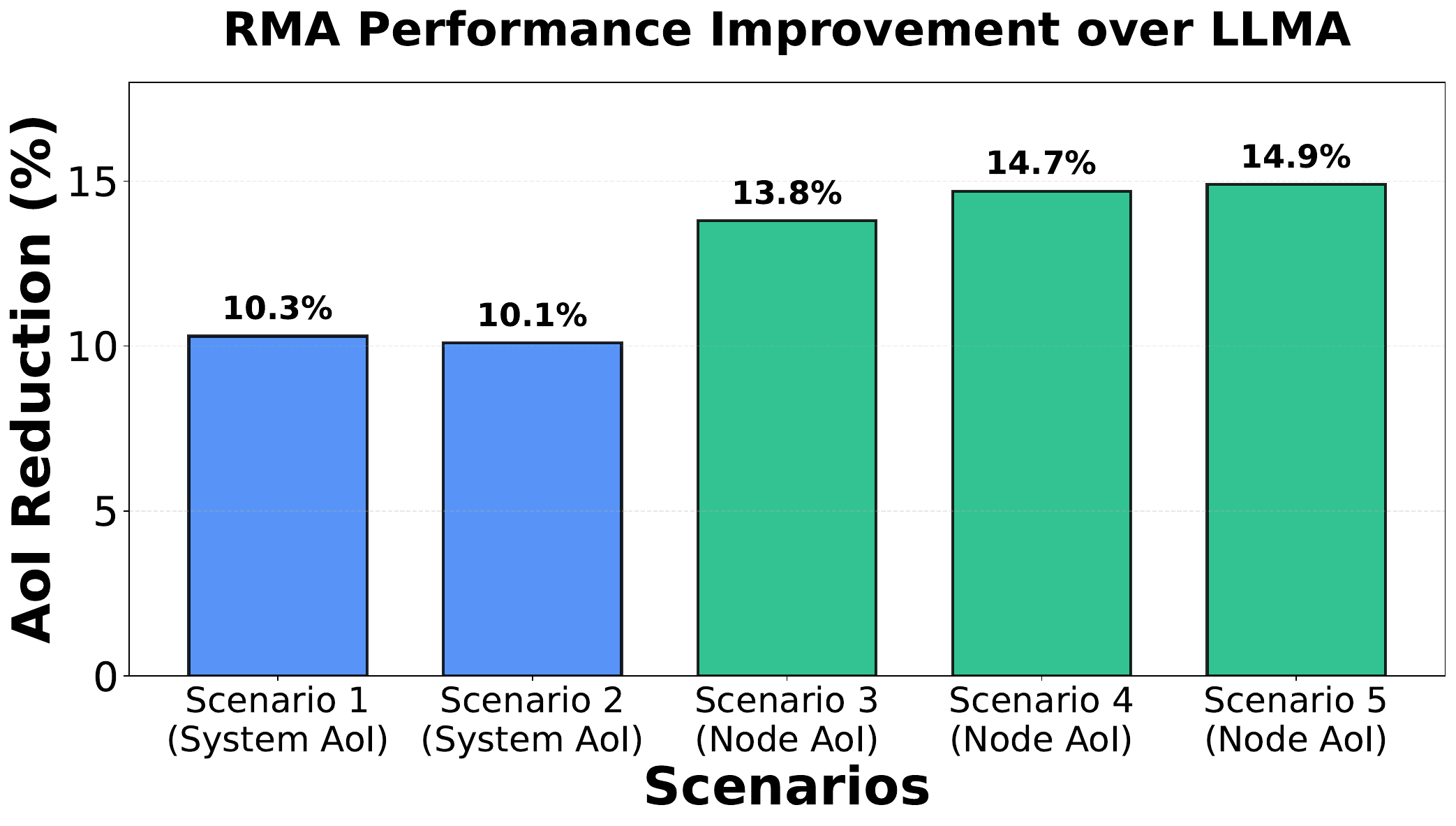}
          \scriptsize{(f) Performance percentage}
      \end{minipage}

      \vspace{0.3cm}

      \caption{Performance comparison across different scenarios. (a)-(e) illustrate the AoI performance of RMA and LLMA methods in Scenarios 1-5, covering both homogeneous (Scenario 3) and heterogeneous (Scenarios 1-2, 4-5) network environments. (f) presents the overall performance percentage improvement, with RMA achieving 10.3-14.9\% AoI reduction compared to LLMA across all scenarios.}
      \label{fig:performance_comparison}
  \end{figure*}
  
\section{Simulations}
In this section, we conduct extensive experiments to evaluate the performance of the proposed Reflex-Core framework and the RMA protocols. We begin by detailing the experimental setups and the design of diverse heterogeneous scenarios. Then, we benchmark RMA against state-of-the-art baselines to demonstrate its superiority in minimizing the system's AoI. Moreover, we investigate the framework's robustness in dynamic network environments and 
its capability in minimizing AoI of the practical systems through a semantic-based priority mechanism. Finally, we present an ablation study to quantify the individual contributions of the Reflection, SFT, and PPO modules in the Reflex-Core framework.

\subsection{Experimental Settings}
\subsubsection{Implementation Details}
All experiments are based on the LongChat-7B-16k model (16k context length and the temperature is $0$) deployed on a laboratory server with NVIDIA A100 GPU (80GB HBM2), AMD EPYC 7742 CPU, and 256GB RAM running Ubuntu 20.04. 
This model has a strong dialogue capability, which is essential for processing long reflection texts and supporting complex reasoning tasks in our framework. In addition, its open-source nature enables convenient local deployment and fine-tuning.
We consider Python 3.8, PyTorch 2.0, and 8-bit quantization for efficient model training and computation.
For training, we collect 500 input-output samples with approximately 1,500 tokens per sample. 

\subsubsection{Baselines}

We consider CP-AgentNet \cite{kwon2025cp}, which is also an LLM-driven framework for autonomous communication protocol design using generative agents, as one of our baselines. While CP-AgentNet employs strategy refinement during its design phase, it lacks the specific reflection mechanism proposed in this paper. The nodes employing CP-AgentNet are referred to as LLMA nodes.

We also compare with DLMA \cite{yu2019deep}, a representative deep reinforcement learning (DRL) based multiple access protocol. It utilizes a deep residual network (ResNet) to learn transmission strategies through numerical state-action-reward interactions. However, it lacks the semantic reasoning and explainability inherent in our LLM-based RMA framework due to its ``black-box" nature.

\subsubsection{Scenario Design}
To comprehensively evaluate the performance of the RMA protocol in different communication environments, we consider the following five scenarios:
\begin{itemize}
    \item \textbf{Scenario 1:} Heterogeneous environment composed of 1 ALOHA and 1 TDMA nodes, and 1 heteronode;
    \item \textbf{Scenario 2:} Heterogeneous environment composed of 3 ALOHA and 1 TDMA nodes, and 1 heteronode;
    \item \textbf{Scenario 3:} Homogeneous environment composed of 2 heteronodes;
    \item \textbf{Scenario 4:} Heterogeneous environment composed of 1 ALOHA and 1 TDMA nodes, and 2 heteronodes;
    \item \textbf{Scenario 5:} Heterogeneous environment composed of 3 ALOHA and 1 TDMA nodes, and 2 heteronodes.
\end{itemize}

Note that we use the term ``heteronode” adhering to \cite{kwon2025cp} to refer to nodes such as the RMA node,
DLMA node, or LLMA node.
For RMA nodes, each observation window contains $N=200$ time slots, where each time slot lasts 1ms. When the number of observation windows within a reflection cycle reaches the preset threshold $O=3$, the Reflect module is triggered to conduct reflection. 
For ALOHA nodes, the transmission probability is set to $q=0.2$; for TDMA nodes, each frame contains 10 time slots, with time slots 3 and 5 fixed for TDMA node transmission. 
We run each experimental scenario for a sufficient duration to ensure convergence.

\subsubsection{Evaluation Metric Settings}
We consider two types of performance metrics for different network scenarios.
  \begin{itemize}
    \item \textbf{Scenario 1-2 (System-level Optimization)}: Use system average AoI as the main evaluation metric, focusing on overall network information timeliness optimization effectiveness.

    \item \textbf{Scenario 3-5 (Node-level Optimization)}: Use heteronode's own average AoI as the main evaluation metric, focusing on individual performance of RMA protocol in competitive environments.
  \end{itemize}

\subsection{Performance Comparison}

\subsubsection{Benchmark Experimental Results}
We compare the AoI performance between LLMA nodes and RMA nodes. In terms of system average AoI optimization (Scenarios 1-2), as shown in Fig.~\ref{fig:performance_comparison}(a)-(b). We see that RMA reduces system average AoI by approximately $10.3\%$ in Scenario 1 and $10.1\%$ in Scenario 2 compared with LLMA, demonstrating consistent performance improvement in heterogeneous environments with varying complexity.
For the average AoI optimization (Scenarios 3-5) case, as shown in Fig.~\ref{fig:performance_comparison}(c)-(e), RMA achieves significant AoI reduction across all scenarios. It can be seen in Scenario 3 that RMA reduces AoI by $13.8\%$; While in the moderate (Scenario 4) and highly competitive (Scenario 5) complex heterogeneous environment, RMA reduces the AoI by $14.7\%$ and $14.9\%$, respectively, compared with LLMA.

Finally, Fig.~\ref{fig:performance_comparison}(f) presents the overall performance percentage improvement across all scenarios. We see that Reflex-Core exhibits a superior performance compared with the CP-AgentNet framework over five scenarios. The performance gap increases when the network scenario becomes more complex.
These results underscore the effectiveness of the proposed reflection mechanism.

\subsubsection{Robustness in Dynamic Scenarios}

\begin{figure}[!t]
    \centering
    \includegraphics[width=0.48\textwidth]{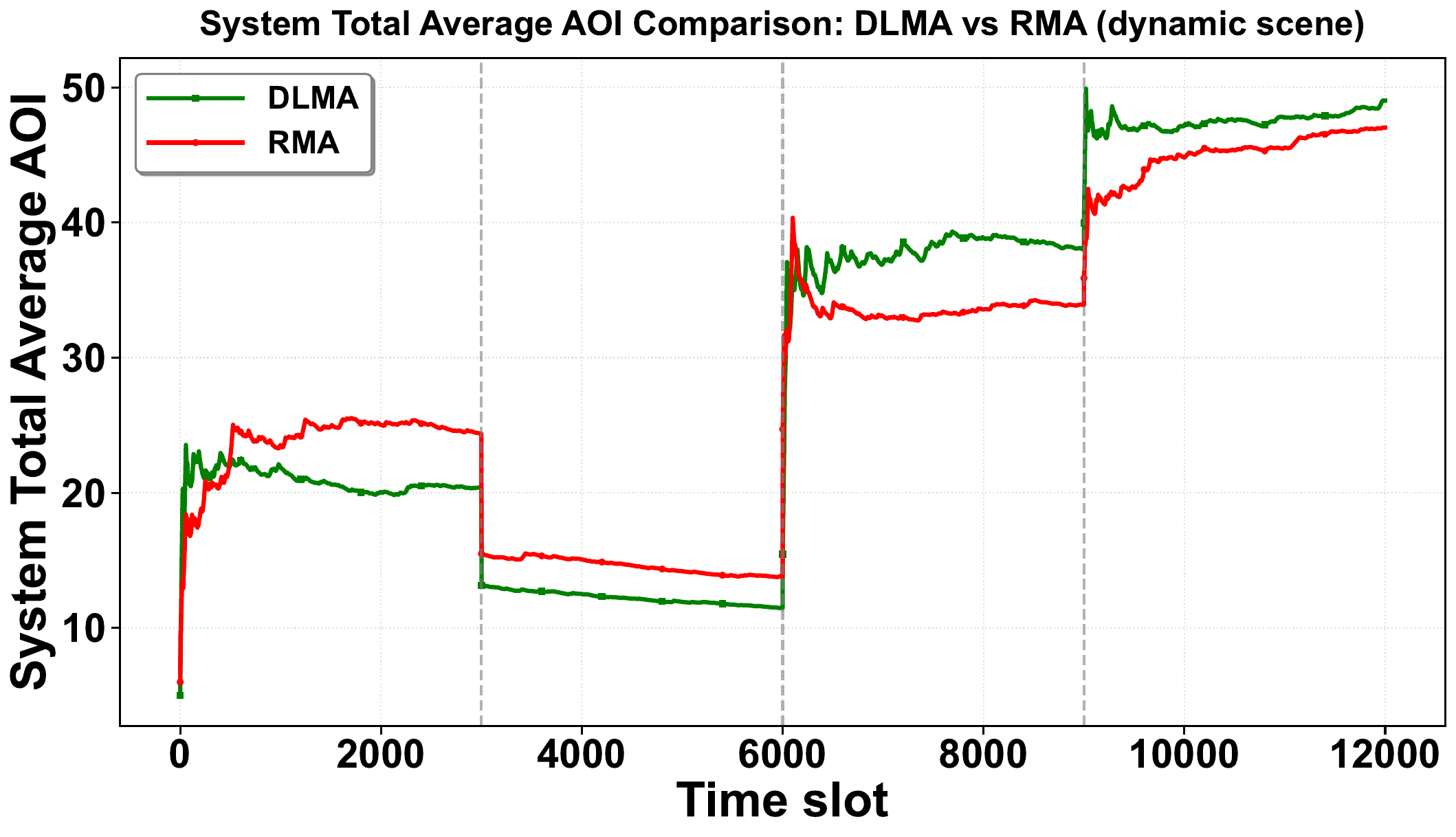}
    \caption{Performance comparison in dynamic scenarios. The network topology changes at $T=3000, 6000, 9000$ slots. Results show that RMA achieves superior robustness and faster re-convergence in high-complexity stages (Stage 3 and 4) compared to DLMA.}
    \label{fig:dynamic}
\end{figure}

We evaluate the adaptability of the RMA protocol in a dynamic setting to assess its responsiveness to environmental changes. As shown in Fig. \ref{fig:dynamic}, this scenario starts with two slotted ALOHA nodes and one heteronode. At slot $3,000$, one ALOHA node exits, followed by the entry of two new ALOHA nodes at slot 6000. The scenario adds a TDMA node at slot $9,000$. 

As illustrated in Fig.~\ref{fig:dynamic}, the results reveal a clear performance inversion as the network complexity increases. In the low-density Stages 1 and 2, the data-driven DLMA achieves a lower average AoI ($20.83$ and $12.23$) compared with RMA ($23.75$ and $14.60$). However, in the high-density and heterogeneous Stages 3 and 4, RMA significantly outperforms DLMA, achieving a $10.76\%$ and $5.34\%$ reduction in average AoI, respectively. 
Such an inversion demonstrates that while DRL methods excel in simplistic tasks, the semantic reasoning of RMA effectively enhances scalability and robustness in non-stationary networks by ``understanding'' underlying protocol logic rather than merely fitting numerical rewards.

\subsubsection{Performance Comparison under Priority Mechanism}
\begin{figure*}[!t]
      \centering
      \begin{minipage}{0.32\textwidth}
          \centering
          \includegraphics[width=\textwidth]{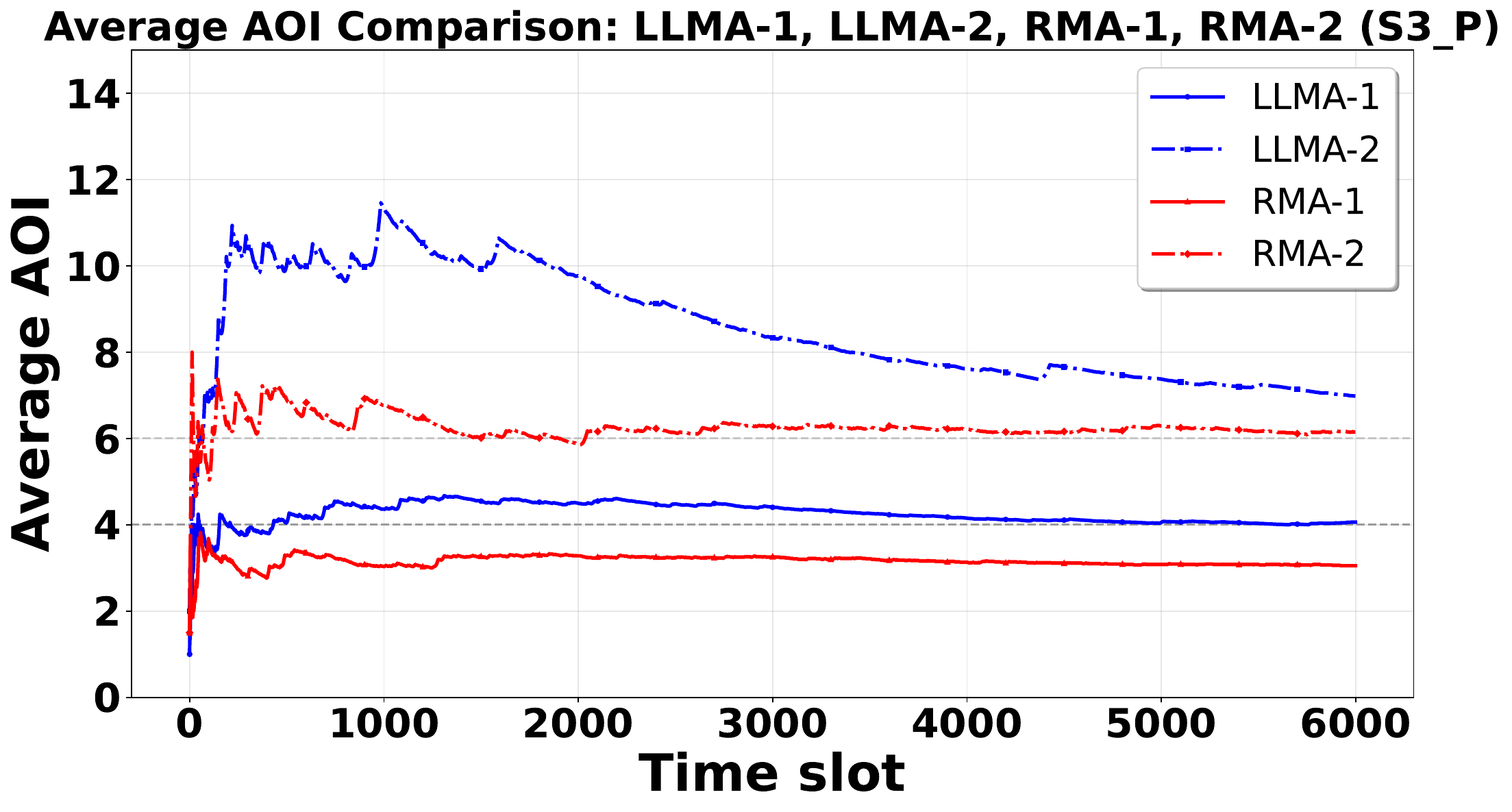}
          \scriptsize{(a) Scenario 3}
      \end{minipage}
      \hfill
      \begin{minipage}{0.32\textwidth}
          \centering
          \includegraphics[width=\textwidth]{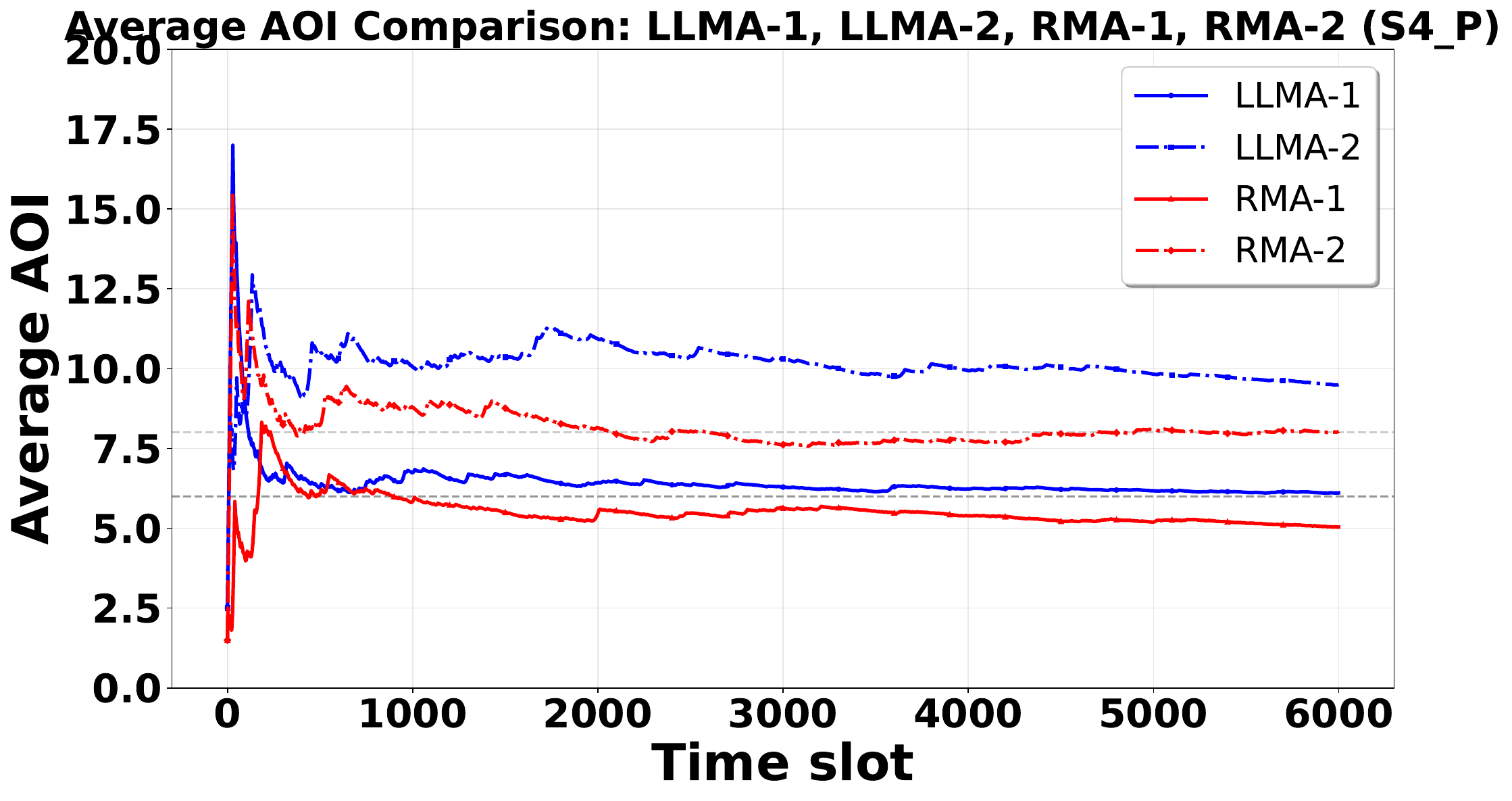}
          \scriptsize{(b) Scenario 4}
      \end{minipage}
      \hfill
      \begin{minipage}{0.32\textwidth}
          \centering
          \includegraphics[width=\textwidth]{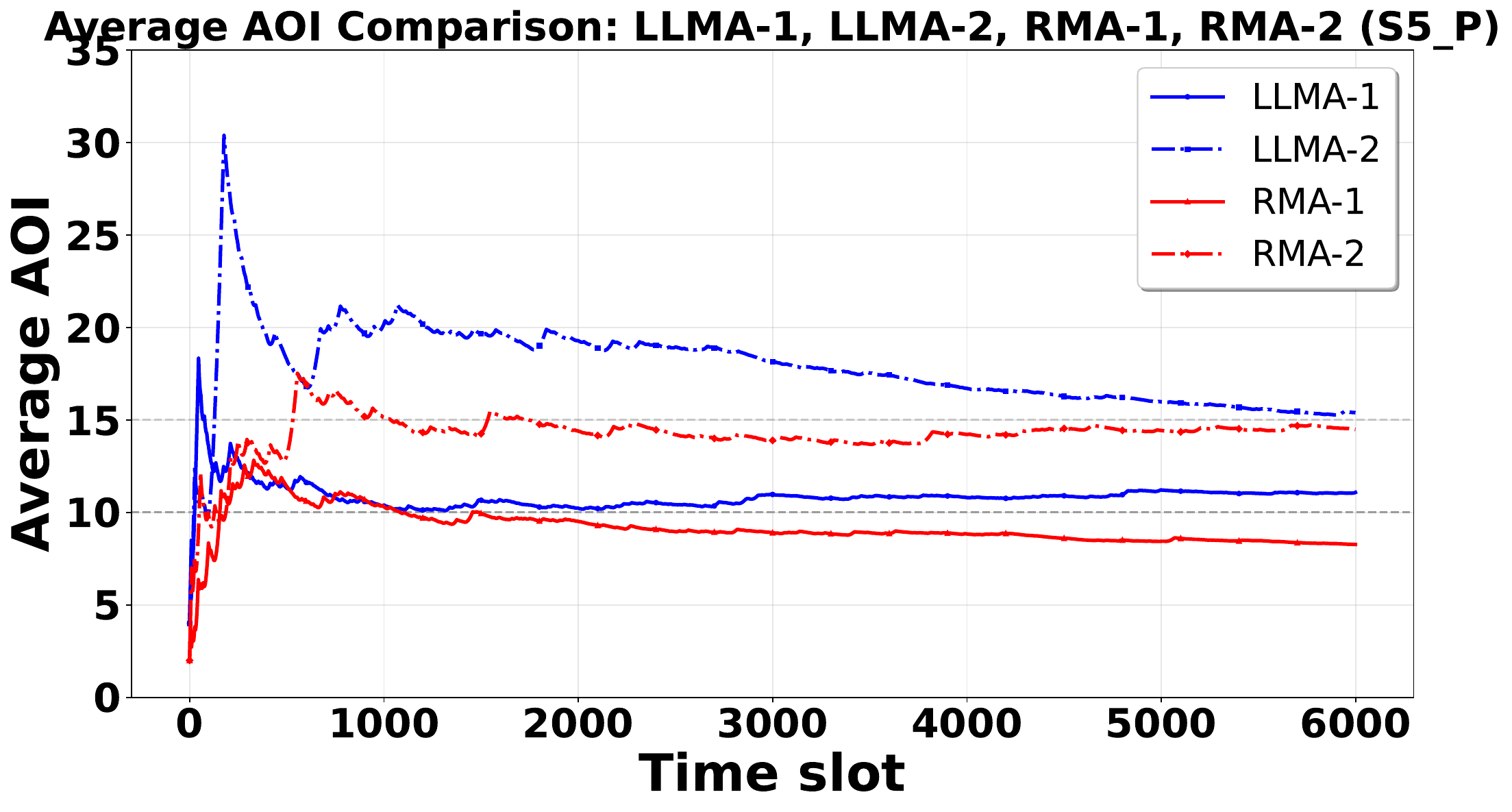}
          \scriptsize{(c) Scenario 5}
      \end{minipage}

      \vspace{0.3cm}

      \caption{Performance comparison under priority mechanism. (a)-(c) illustrate the AoI performance of RMA and LLMA methods in Scenarios 3-5 with priority scheduling. Results demonstrate that RMA maintains significant optimization advantages, achieving $14.2-17.2\%$ AoI reduction compared to LLMA across different priority environments.}
      \label{fig:priority_comparison}
  \end{figure*}

To further validate the effectiveness of the priority mechanism, we compare the performance of LLMA and RMA methods in priority environments based on Scenarios 3-5, as shown in Fig.~\ref{fig:priority_comparison}(a)-(c). We introduce differentiated target AoI thresholds ($\delta_{th}$) to characterize the Quality of Service (QoS) requirements for the high-priority (HP) and low-priority (LP) nodes. Specifically, the $(\text{HP, LP})$ threshold pairs are set to $(4, 6)$ for Scenario 3, $(6, 8)$ for Scenario 4, and $(10, 15)$ for Scenario 5, which are visualized as horizontal gray dashed lines. 
These thresholds serve as differentiated QoS benchmarks: for HP nodes, the primary objective is to rapidly drive the AoI below $\delta_{th}$ to ensure superior information freshness, whereas LP nodes aim to maintain stability near their respective thresholds.

   The results demonstrate that under the priority mechanism, the RMA method still significantly outperforms the LLMA method in terms of:

  \begin{itemize}
      \item \textbf{Average Node AoI Reduction Effect}: After introducing the priority mechanism, RMA successfully achieves a significant reduction in average node AoI compared to LLMA. In the homogeneous environment of Scenario 3 (Fig.~\ref{fig:priority_comparison}(a)), average node AoI is reduced by approximately $16.3\%$; in the moderately complex heterogeneous environment of Scenario 4 (Fig.~\ref{fig:priority_comparison}(b)), average node AoI is reduced by approximately $17.2\%$; in the highly complex heterogeneous environment of Scenario 5 (Fig.~\ref{fig:priority_comparison}(c)), average node AoI is reduced by approximately $14.2\%$. These results show that RMA's reflection mechanism maintains its optimization advantages in priority environments.

      \item \textbf{Threshold Arrival Speed Improvement}: Experiments observed that nodes in the RMA method reach preset AoI thresholds faster than in the LLMA method. As shown in Fig.~\ref{fig:priority_comparison}(a)-(c), experimental results show that RMA improves node threshold arrival speed by approximately $15-20\%$ compared to LLMA, indicating that the RMA method has  advantages in convergence rate.
  \end{itemize}


  Through the above comparison results, the following conclusions can be drawn.

  \begin{itemize}
    \item \textbf{Reflection Mechanism Advantages}: Whether introducing priorities or not, Reflex-Core outperforms CP-AgentNet, validating the effectiveness and universality of the reflection learning framework.

    \item \textbf{Priority Mechanism Effectiveness}: Priority scheduling successfully achieves differentiated QoS guarantees, validating the effectiveness of Reflex-Core in handling multi-priority requirements.

    \item \textbf{Dynamic Scenario Adaptability}: Experimental results show that Reflex-Core  maintains  performance gains within complex, dynamic communication environments, proving the method's robustness and practical utility.
  \end{itemize}

\subsection{Ablation Study}
In the following, we present an ablation study to evaluate the effectiveness of each component in the Reflex-Core framework by 
considering four configurations: (1) Single-LLM as the baseline; (2) RMA without reflection; (3) RMA with reflection and reflection-oriented SFT; and (4) Full RMA with the complete training pipeline (SFT + reward modeling + PPO).
The experiment is conducted across three representative scenarios with varying traffic conditions and protocol mixes: Scenario 1 (moderate load, heterogeneous protocols), Scenario 2 (high load), and Scenario 3 (node-level AoI optimization focus).

\begin{figure*}[!t]
    \centering
    \begin{minipage}{0.32\textwidth}
        \centering
        \includegraphics[width=\textwidth]{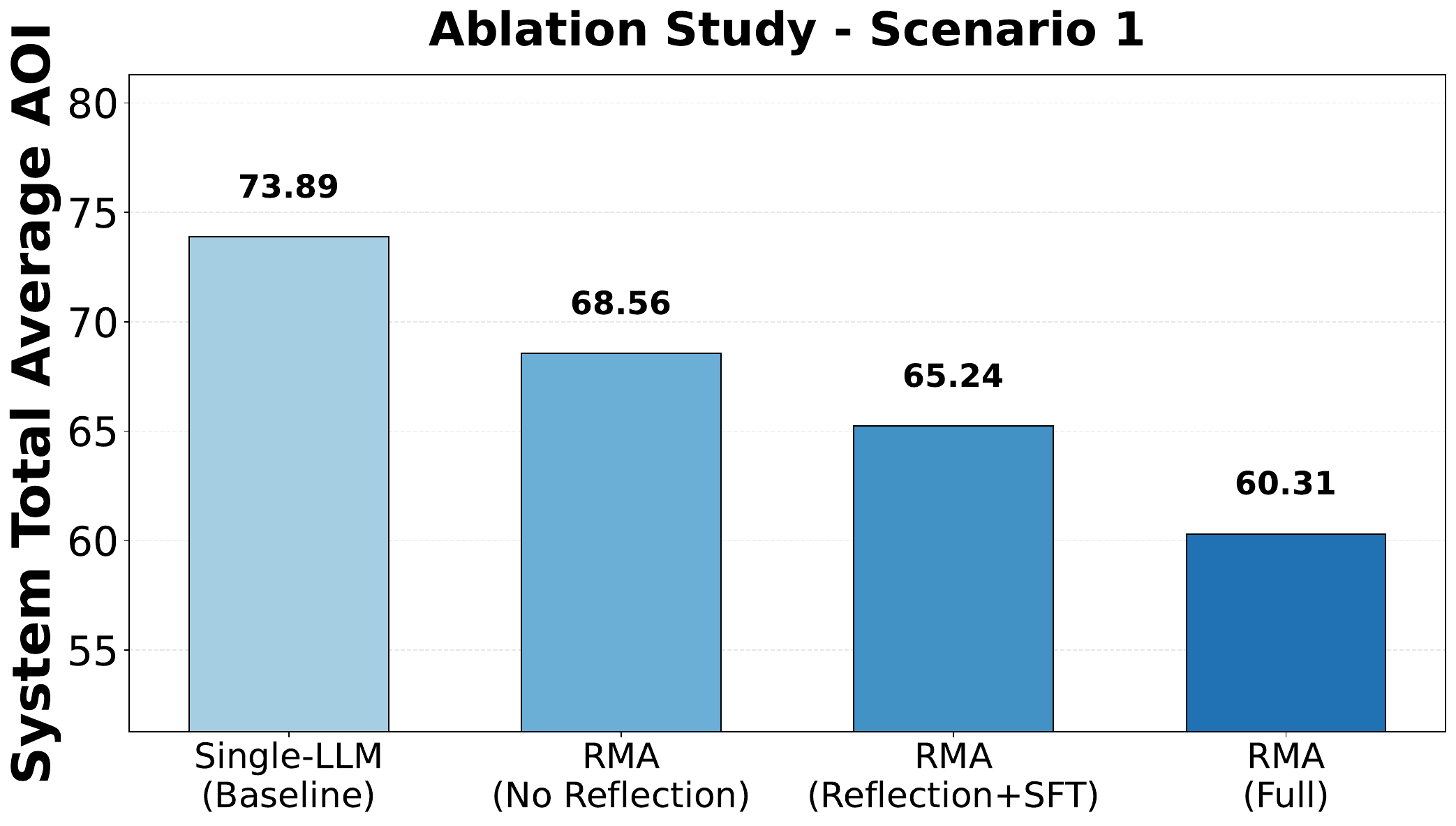}
        \scriptsize{(a) Scenario 1}
    \end{minipage}
    \hfill
    \begin{minipage}{0.32\textwidth}
        \centering
        \includegraphics[width=\textwidth]{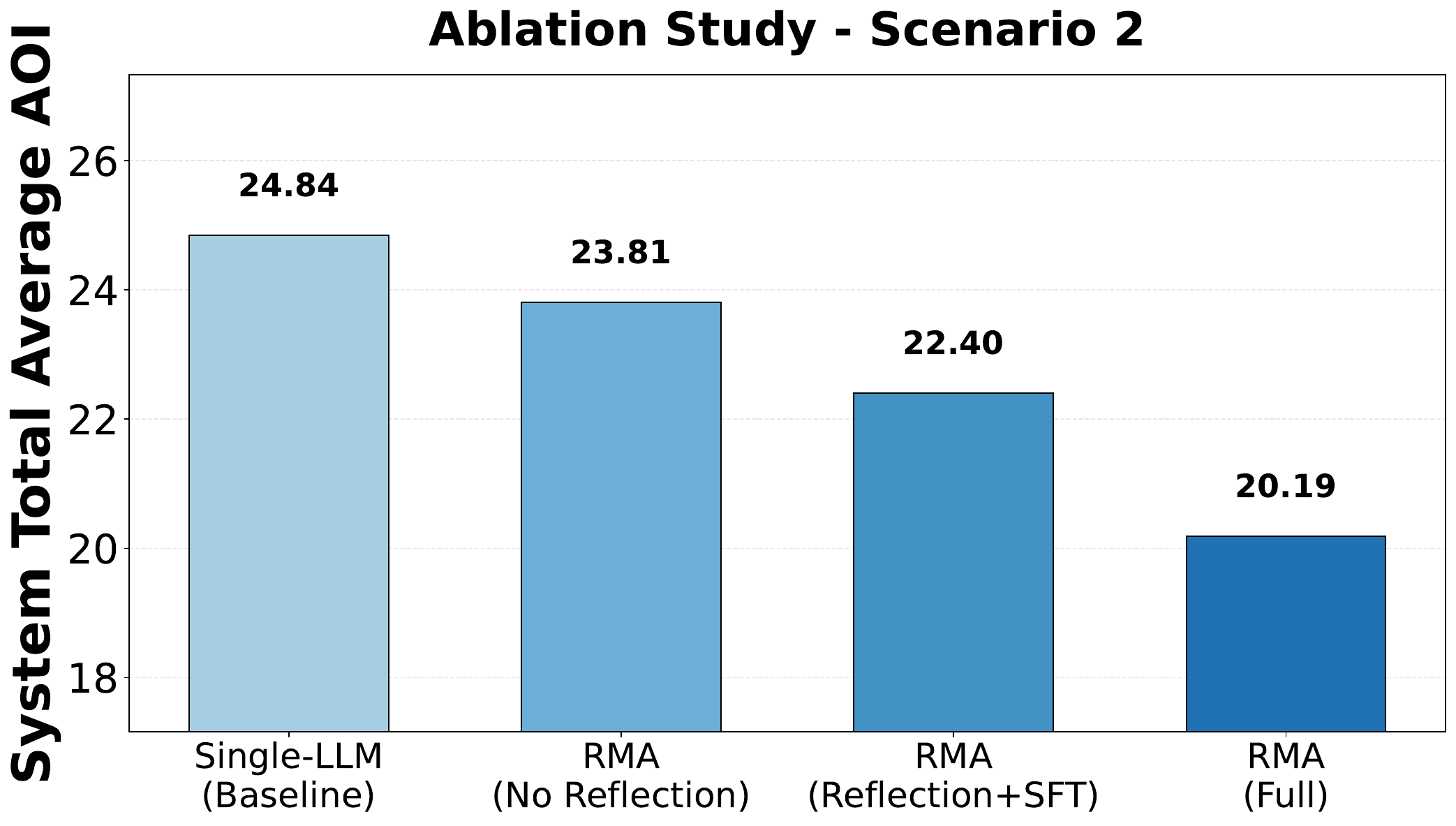}
        \scriptsize{(b) Scenario 2}
    \end{minipage}
    \hfill
    \begin{minipage}{0.32\textwidth}
        \centering
        \includegraphics[width=\textwidth]{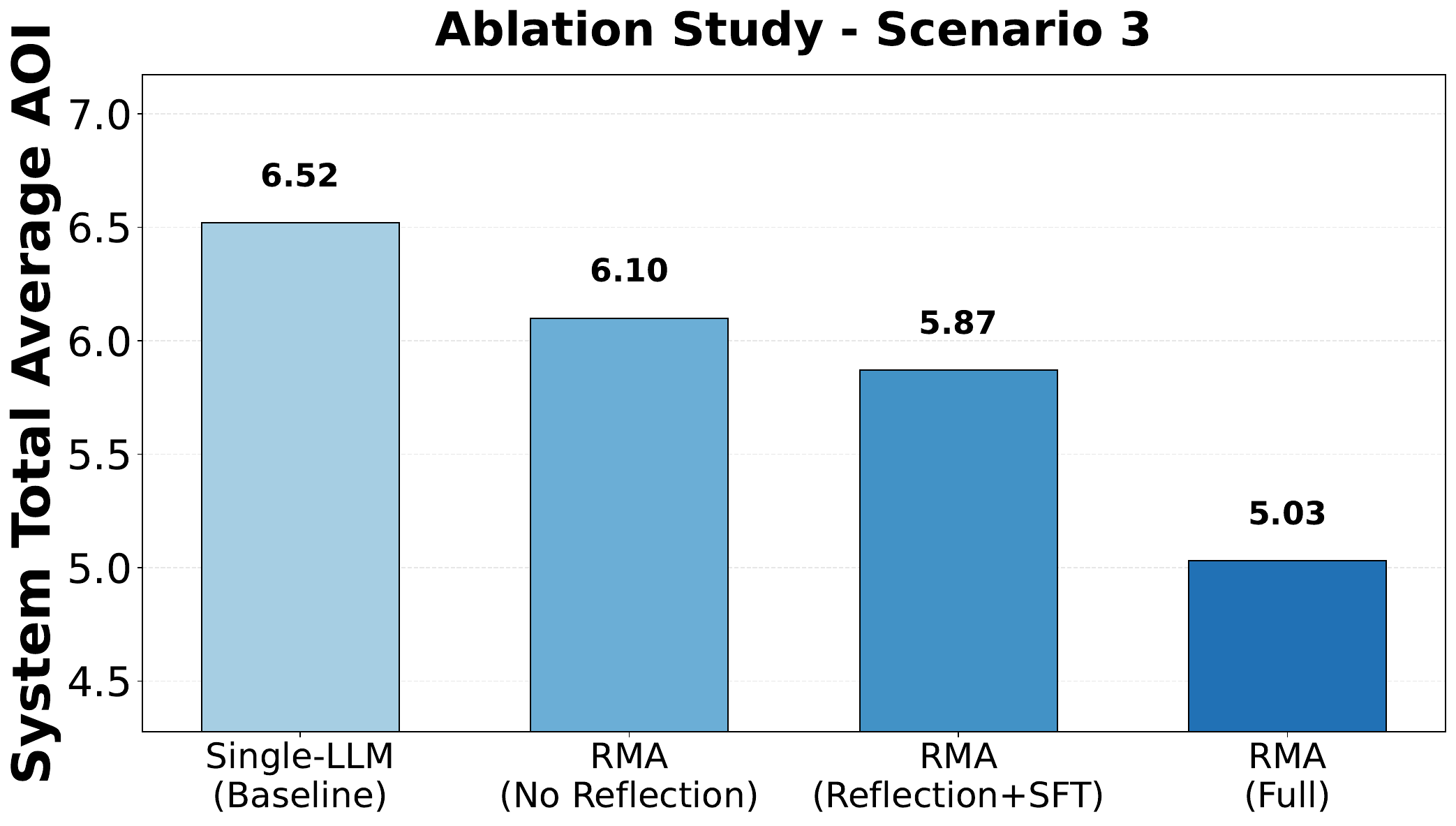}
        \scriptsize{(c) Scenario 3}
    \end{minipage}

    \vspace{0.2cm}
    \caption{Ablation study results across three scenarios. (a)-(c) show the system total average AoI for Single-LLM baseline, RMA without reflection, RMA with reflection+SFT, and full RMA system. Results demonstrate that each component contributes incrementally: reflection provides $4-5\%$ improvement, the complete training pipeline adds $6-7\%$ gains, and the full RMA system achieves $18-23\%$ total AoI reduction compared to the baseline.}
    \label{fig:ablation}
\end{figure*}

Fig.~\ref{fig:ablation} shows consistent improvements across all scenarios. In Scenario 1, starting from the Single-LLM baseline (AoI is $73.89$), RMA without reflection achieves a $7.21\%$ reduction (AoI is $68.56$). Adding reflection and SFT further improves the performance to an $11.71\%$ reduction (AoI is $65.24$). The full RMA system achieves the best performance with an 18.38\% total reduction (AoI is 60.31). Similar trends are observed in Scenarios 2 and 3, where full RMA achieves $18.72\%$ ($20.19$ vs. baseline $24.84$) and $22.85\%$ ($5.03$ vs. baseline $6.52$) reductions, respectively.

These consistent performance improvements underscore the necessity of a closed-loop cognitive architecture for complex network control. Indeed, the leap from a basic LLM to a reflective agent suggests that the ability to semantically diagnose past failures is key to navigating non-stationary interference patterns. Moreover, the synergy between SFT and PPO-based training confirms that domain-specific alignment is essential; it transforms the LLM from a general-purpose reasoner into a specialized network controller that can map linguistic insights onto high-performance transmission strategies. Therefore, these results demonstrate that the full RMA system achieves its superior information freshness not through random exploration, but through a structured evolution of policy refinement facilitated by its internal modular components. 

\section{Conclusion}
In this paper, we proposed \emph{Reflex-Core}, an interpretable LLM-agent framework for AoI optimization. Through a novel ``Observe-Reflect-Decide-Execute'' closed-loop mechanism, this framework facilitates semantic-level decision-making in heterogeneous random access networks. Based on Reflex-Core, we devised and implemented the RMA protocol by integrating the SFT and PPO techniques, alongside a priority-based RMA protocol to support differentiated QoS requirements. Experiment results demonstrated that Reflex-Core significantly outperforms existing baselines, particularly in complex, dynamic environments.

Future work will focus on improving the computational efficiency for edge devices, extending the framework to multi-objective scenarios, and ensuring robustness in adversarial settings. Ultimately, this study offers a new paradigm for random access control, highlighting the potential of LLM-based intelligent optimization in next-generation communication systems.

\appendix
\subsection*{A. Normal Mode Prompts}
  \label{appendix:normal}

  \begin{tcolorbox}[breakable, title={Observe Module Prompt}]
  \textbf{}\\
  \footnotesize
  \textbf{Prompt:} As the Observe module of the communication protocol system, your core task is to analyze the execution effect of strategies within the current observation window and generate strategy adjustment suggestions for the next observation window, with the goal of
  minimizing the system's average AoI.\\

  In the observation analysis process, you will receive the following information:\\
  1. Window execution strategy: The actual transmission probability used in the current window $p' = \texttt{\{strategy\_executing\}}$\\
  2. Current cycle strategy: The global transmission strategy set in the previous reflection cycle $p = \texttt{\{strategy\_output\}}$\\
  3. Current window system average AoI: The AoI change sequence within the current observation period $\texttt{\{aoi\_current\}}$, reflecting the AoI change trend within this window\\
  4. Previous window system average AoI: The AoI change sequence within the previous observation period $\texttt{\{aoi\_previous\}}$, used for comparative analysis\\
  5. Current window state distribution: Channel state statistics for the current observation period $\texttt{\{states\_current\}}$, containing the occurrence counts of states such as success, collision, other nodes successful when not transmitting, other nodes collided when
  not transmitting, and system idle\\
  6. Previous window state distribution: Channel state statistics for the previous observation period $\texttt{\{states\_previous\}}$, containing the same state classification statistics\\

  Based on this information, please conduct observation and analysis from two dimensions:\\
  1. Performance trend analysis: Compare the AoI changes between the current window and the previous window to analyze the improvement or deterioration trend of system performance\\
  2. Strategy adjustment suggestions: Based on the channel state distribution and AoI changes, propose specific transmission probability adjustment suggestions for the next window\\
  \\
  \textbf{Output:} 1. Performance trend analysis
  The current window AoI of 3.25 shows improvement compared to the previous window's 3.45, and the effect is significant after adjusting the strategy from 0.40 to 0.35. The state distribution shows that the collision rate decreased from 20\% to 7.5\%, indicating that the
  strategy adjustment effectively reduced conflicts, but the idle rate increased from 50\% to 60.5\%, suggesting it may be too conservative, with insufficient channel utilization.

  2. Strategy adjustment suggestions
  Based on the analysis of reduced collision rate but increased idle rate, it is recommended to slightly increase the transmission probability by 0.02 to better utilize channel opportunities and further improve information freshness. The current optimization direction is
  correct, and appropriately increasing transmission frequency can improve transmission efficiency while maintaining low collision rates.
  \\
  \end{tcolorbox}

  \begin{tcolorbox}[breakable, title={Reflect Module Prompt}]
  \textbf{}\\
  \footnotesize
  \textbf{Prompt:} As the Reflect module of the communication protocol system, your core task is to analyze the execution effects of strategies and generate optimization suggestions, with the goal of minimizing the system's average AoI.\\

  During the analysis process, you will receive the following information:\\
  1. Long-term memory: Performance summary of all observation windows within this reflection cycle $\texttt{\{long\_term\_memory\}}$, including AoI change sequences, execution strategies, and state distribution statistics (occurrence counts of success, collision, other nodes
  successful when not transmitting, other nodes collided when not transmitting, system idle, etc.) for each window\\
  2. Current cycle strategy: The global transmission strategy set in the previous reflection cycle $p = \texttt{\{strategy\_output\}}$\\
  3. System average AoI difference: The change in system average AoI between this reflection cycle and the previous reflection cycle $\texttt{aoi\_delta} = \texttt{\{aoi\_delta\}}$ (negative value indicates AoI reduction, which is good improvement)\\
  4. Previous reflection feedback: $\texttt{\{last\_reflection\}}$ the previous round's reflection analysis record, containing strategy problem diagnosis and improvement suggestions\\

  Based on this information, please conduct in-depth reflection and analysis from two dimensions:\\
  1. Strategy effect evaluation: Analyze the current strategy execution performance, verify the effect of the last reflection strategy adjustment, identify existing problems and optimization directions\\
  2. Improvement plan design: Based on the previous strategy effect evaluation and combined with historical reflection experience, propose specific numerical adjustment suggestions for strategy optimization\\
  \\
  \textbf{Output:}
  1. Strategy effect analysis: Comprehensive analysis within the reflection cycle shows that the transmission strategy $p=0.35$ performs overall well in this round, validating the correctness of the previous reflection strategy adjustment. Specific performance: the average AoI
  of this reflection cycle is 45.1 compared to the previous 48.7, with $\text{aoi\_delta}=-3.6$ indicating the correct direction of strategy adjustment. Reviewing the previous reflection's suggestions, it was found at that time that the strategy was too aggressive leading to
  high collision rates, so it was recommended to reduce the transmission probability from 0.40 to 0.35. This adjustment was validated in this round, with the collision rate decreasing from 15\% to 7.4\%. However, current observations suggest the strategy may have been
  adjusted excessively, with significant channel underutilization: the average successful transmission rate is only 9.6\%, while the proportion of other nodes succeeding reaches 21\%, indicating the current strategy is too conservative, missing numerous available transmission
   opportunities and affecting further AoI improvement.\\
  2. Improvement plan design:
  It is recommended to increase the transmission probability by 0.03 to $p=0.38$. This adjustment can improve the channel underutilization issue while remaining within the experience range of similar strategies in the strategy memory. The current optimization direction is
  correct, and appropriately increasing transmission frequency can improve transmission efficiency while maintaining low collision rates.\\
  \end{tcolorbox}

  \begin{tcolorbox}[breakable, title={Decide Module Prompt}]
  \textbf{}\\
  \footnotesize
  \textbf{Prompt:} As the Decide module in the communication protocol system, your core task is to update the system's global transmission strategy based on the Reflect module's optimization suggestions and maintain the strategy memory base, with the goal of minimizing the
  system's average AoI. Memory-based management is based on the actual AoI improvement effect of strategies for storage decisions.\\

  During the strategy formulation process, you will receive the following information:\\
  1. Current reflection feedback: $\texttt{\{current\_reflection\}}$ Strategy analysis report and optimization suggestions generated by the Reflect Agent\\
  2. Current cycle strategy: The global transmission strategy set in the previous reflection cycle $p = \texttt{\{strategy\_output\}}$\\
  3. Strategy memory: $\texttt{\{strategy\_memory\}}$ Historical optimal strategy set, strategy records screened and saved based on AoI improvement effect\\
  4. System average AoI difference: The change in system average AoI between this reflection cycle and the previous reflection cycle $\texttt{aoi\_delta} = \texttt{\{aoi\_delta\}}$ (negative value indicates AoI reduction, which is good improvement)\\

  Based on this information, please conduct strategy analysis from two dimensions:\\
  1. Strategy effect evaluation: Analyze the historical performance of the current strategy, combine current reflection feedback and strategy memory, evaluate strategy effectiveness and predict new strategy effects\\
  2. Strategy generation and memory management: Generate new global transmission strategies, while deciding whether to store the current strategy in the strategy memory base based on actual AoI improvement effects\\

  \textbf{Output:}
  1. Strategy effect evaluation: The current strategy $p=0.35$ performed well in the previous reflection cycle, with $\text{system\_delta}=-2.8$ indicating the correct direction of strategy adjustment, effectively reducing system AoI. Reflect module analysis confirms that the
  current strategy has channel under-utilization issues with room for further optimization. In the strategy memory, $p=0.38$ performed excellently in similar environments, providing a reference for the current adjustment.\\
  2. Strategy generation and memory management:
  Based on the Reflect module's suggestion, increase the transmission probability by 0.03 to $p=0.38$. This adjustment can improve the channel under-utilization issue while remaining within the experience range of similar strategies in the strategy memory. Since the current
  strategy $p=0.35$ brought significant AoI improvement ($\text{system\_delta}=-2.8$), it is recommended to store it in the strategy memory base as a historically successful strategy. Strategy output: [$p=0.38$]
  Memory update: Store current strategy $p=0.35$ in strategy memory base\\
  \end{tcolorbox}

  \subsection*{B. Priority Mode Prompts}
  \label{appendix:priority}

  \begin{tcolorbox}[breakable, title={Reflect Module Prompt (Priority Mode)}]
  \textbf{}\\
  \footnotesize
  \textbf{Prompt:} As the Reflect module of the communication protocol system, your core task is to analyze the execution effects of strategies and generate optimization suggestions, with the goal of minimizing the node's average AoI. During the analysis process,
  you need to adjust the analysis strategy according to the system operating mode: in priority mode, high-priority nodes (RMA1) should adopt more aggressive analysis methods to fully utilize priority advantages for greater performance improvements; low-priority nodes (RMA2)
  need to maintain a balance between performance optimization and system coordination, adopting moderate strategy adjustments; in normal mode, all nodes have equal status and are unified for AoI optimization analysis.\\

  ... [Same input information as Normal Mode] ...

  Based on this information, please conduct in-depth reflection and analysis from two dimensions:\\
  1. Strategy effect evaluation: ... [Same as Normal Mode]\\
  2. Improvement plan design: Based on the previous strategy effect evaluation and combined with historical reflection experience, propose specific numerical adjustment suggestions according to node priorities (unified adjustment in normal mode, more aggressive for
  high-priority nodes RMA1 in priority mode, more moderate for low-priority nodes RMA2)\\
  \\
  \textbf{Output:}
  1. Strategy effect analysis: ... [Same as Normal Mode]\\
  2. Improvement plan design:
      - Normal mode: It is recommended to increase the transmission probability by 0.03 to $p=0.38$.
      - RMA1 high priority: Based on the advantages and current performance of high-priority nodes, it is recommended to significantly increase the transmission probability by 0.06 to $p=0.41$ to fully utilize priority advantages for rapid AoI reduction.
      - RMA2 low priority: Noting that the AoI of high-priority nodes has not yet reached the threshold, it is recommended to slightly adjust the transmission probability by 0.02 to $p=0.37$ to modestly improve performance while yielding to high-priority nodes.\\
  \end{tcolorbox}

  \begin{tcolorbox}[breakable, title={Decide Module Prompt (Priority Mode)}]
  \textbf{}\\
  \footnotesize
  \textbf{Prompt:} As the Decide module in the communication protocol system, your core task is to update the system's global transmission strategy based on the Reflect module's optimization suggestions and maintain the strategy memory base, with the goal of minimizing the
  node's average AoI. During the strategy generation process, you need to adjust the strategy method according to the system mode: in priority mode, high-priority nodes (RMA1) should more actively adopt aggressive suggestions to fully utilize priority advantages for strategy
  adjustment; low-priority nodes (RMA2) need to moderately adopt suggestions to maintain a balance between performance improvement and system coordination; in normal mode, all nodes have equal status and are unified for strategy adjustment. Memory base management is based on
  the actual AoI improvement effect of strategies for storage decisions.\\

  ... [Same input information as Normal Mode] ...

  Based on this information, please conduct strategy analysis from two dimensions:\\
  1. Strategy effect evaluation: ... [Same as Normal Mode]\\
  2. Strategy generation and memory management: ... [Same as Normal Mode]\\

  \textbf{Output:}
  1. Strategy effect evaluation: ... [Same as Normal Mode]\\
  2. Strategy generation and memory management:
      - Based on the Reflect module's (normal mode) suggestion, increase the transmission probability by 0.03 to $p=0.38$. This adjustment can improve the channel underutilization issue while remaining within the experience range of similar strategies in the strategy memory.
  Since the current strategy $p=0.35$ brought significant AoI improvement ($\text{system\_delta}=-2.8$), it is recommended to store it in the strategy memory base as a historically successful strategy. Strategy output: [$p=0.38$]
      Memory update: Store current strategy $p=0.35$ in strategy memory base
      - RMA1 high priority: As a high-priority node, based on the Reflect module's (RMA1 high priority) suggestion, increase the transmission probability by 0.07 to $p=0.42$ to fully utilize priority advantages. Strategy output: [$p=0.41$] Memory update: Store current strategy
  $p=0.35$ in strategy memory base
      - RMA2 low priority: As a low-priority node, based on the Reflect module's (RMA2 low priority) suggestion, maintain a balance between performance improvement and system coordination by increasing the transmission probability by 0.01 to $p=0.36$. Strategy output: [p=0.36]
      Memory update: Store current strategy $p=0.35$ in strategy memory base
  \end{tcolorbox}